\documentclass[aps,prd,twocolumn,groupedaddress]{revtex4-2}
\usepackage{placeins} 
\usepackage{xcolor}
\usepackage{threeparttable}
\usepackage{booktabs} 
\usepackage{bm} 
\usepackage{titlesec} 
\usepackage{amsmath}
\usepackage{graphicx}
\usepackage{caption}
\usepackage{subcaption}
\usepackage{mathtools}
\usepackage{array}
\usepackage{multirow}
\usepackage{amsmath}
\usepackage{colortbl}
\usepackage[colorlinks=true, linkcolor=blue, citecolor=blue, urlcolor=blue]{hyperref}
\begin{document}
\title{Frequency Differences between Clocks on the Earth and the Moon}
\author{Mingyue Zhang}
\author{Jürgen Müller}
\affiliation{Institute of Geodesy (IfE), Leibniz University Hannover, Schneiderberg 50, 30167 Hannover, Germany}
\author{Sergei M. Kopeikin}
\affiliation{Department of Physics \& Astronomy, University of Missouri, 322 Physics Bldg., Columbia, Missouri 65211, USA}
%\date{\today}
\begin{abstract}
Future lunar exploration, navigation, geodesy, and science will require precise comparisons of clocks located on the Earth and the Moon. Motivated by recent advances in optical clock technology reaching fractional frequency uncertainties below $10^{-18}$, we investigate the relativistic frequency-transfer observable between terrestrial and lunar clocks within the framework of the IAU relativistic reference systems.
The frequency comparison is modeled through four time transformations: the proper-to-coordinate time transformations for clocks on the Earth and Moon, the coordinate-time transformation between the lunar and terrestrial reference systems, and the signal-propagation time transfer between the two clocks. The resulting frequency observable is decomposed into contributions arising from the gravity potentials of the Earth and Moon, the transformation between terrestrial and lunar coordinate times, and signal-propagation effects including Doppler, Shapiro, and tropospheric terms.
The above contributions are computed using contemporary Earth and lunar gravity-field models, planetary ephemerides, Earth orientation parameters, and lunar orientation models. Numerical simulations are performed for existing lunar laser-ranging reflectors and representative terrestrial observatories.
We demonstrate that the gravitational-redshift and coordinate-time-transformation terms contribute at the $10^{-10}$--$10^{-11}$ level in fractional frequency, whereas the raw one-way frequency-transfer observable is dominated by first-order Doppler effects at the $10^{-6}$ level. We further show that multi-link Doppler-cancellation schemes suppress the dominant signal-propagation contribution by several orders of magnitude, allowing the underlying relativistic clock-rate signal to be recovered.
The analysis also quantifies the impact of uncertainties associated with gravity-field models, Earth and lunar orientation parameters, station coordinates, and planetary ephemerides on prospective Earth--Moon clock comparisons. Our results provide a relativistic framework for establishing and maintaining a lunar time scale and for supporting future Earth--Moon clock comparisons using next-generation optical clocks.
\end{abstract}
\maketitle
%%%%%%%%%%%%%section-Introduction%%%%%%%%%%%%%%%%%%%
\section{Introduction}
According to general relativity, clock frequency measurements can be used to determine gravity potential \cite{Bjerhammar1985,Bjerhammar1986}. This concept, known as chronometric geodesy, supports various terrestrial studies, e.g., geopotential investigations \cite{Voigt2016,Schroeder2021,Vincent2024a}, sea level monitoring \cite{Vincent2024b}, and height system unification \cite{Wu2018,Wu2022,Shanker2025}. Future lunar clocks could enable some analogous studies on the Moon and integrate with ground and satellite clocks for terrestrial clock comparisons. Under suitable conditions, such as an accurately determined position and local gravity potential at the clock site, adequate environmental control, clock autonomy, and sufficiently accurate Earth--Moon time transfer, a lunar clock could in principle serve as a reference. The Moon's stable orbit, low-disturbance environment, wide Earth visibility, and precisely tracked motion by lunar laser ranging (LLR) provide additional favorable conditions for such a role.

In parallel, clock-based timekeeping underpins future lunar navigation. Projects like ESA’s Moonlight \cite{Giordano2021,Giordano2022,MoonlightTeam2022,MoonlightTeam2022a, Gotta2026} and NASA’s LCRNS \cite{Esper2025, Speciale2022,Ryden2025}, together with the broader LunaNet interoperability framework \cite{NASA2025}, highlight the need for a dedicated lunar time scale \cite{Gibney2023,IAU2024,IAU2024a}, prompting research on its definition, possible realizations and relation to terrestrial standards \cite{Ashby2024,Kopeikin2024,Turyshev2025, Bourgoin2025}.

To support future lunar exploration and precision timekeeping, we develop a relativistic framework for comparing clocks on the Earth and the Moon. The framework combines gravity-potential models, transformations between terrestrial and lunar coordinate times, and signal-propagation effects into a unified description of the observable fractional frequency difference. Unlike previous investigations restricted to terrestrial chronometric geodesy or theoretical analyses of lunar time scales, the present work considers realistic Earth--Moon clock-comparison configurations involving surface-based clocks on both bodies. This extends the application of relativistic frequency metrology and reference-system theory to the cislunar environment and provides an operational basis for future lunar timekeeping.
%%%%%%%%%%%%%section-Proper time and coordinate time%%%%%%%%%%%%%%%%%%%
\section {Simulation method and settings}
Proper time is the time measured by a clock in its own rest frame, while coordinate time is a variable used in a specific reference system to describe the temporal ordering of events \cite{Denker2018,Mueller2018}. The Geocentric Celestial Reference System (GCRS) is a geocentric relativistic reference system that is locally kinematically non-rotating. Its coordinate time is Geocentric Coordinate Time ($u\equiv{\rm TCG}$) \cite{Petit2010}. Analogously, a 2024 resolution of the XXXII IAU General Assembly \cite{IAU2024} defined the Lunar Celestial Reference System (LCRS) and its time counterpart, Lunar Coordinate Time ($s\equiv{\rm TCL}$) for the Moon.

For a clock at rest on the Earth's surface (E-clock $e$), with the gravity potential (Sec.~\ref{sec:grav-pot}) $W_{e}$, the relation between its proper time $\tau_{e}$ and TCG is \cite{Mueller2018} 
\begin{equation}
\frac{d\tau_{e}}{du_{e}} = 1 - \frac{W_{e}}{c^2} + O(c^{-4}),
\label{eq1}
\end{equation}
where $c$ is the speed of light, and $u_{e}$ denotes TCG at the E-clock $e$. 

Similarly, for a clock at rest on the lunar surface (L-clock $\ell$), with the gravity potential $W_{\ell}$, its proper time $\tau_{\ell}$ and TCL are related by \cite{Kopeikin2024}
\begin{equation}
\frac{d\tau_{\ell}}{ds_{\ell}} = 1 - \frac{W_{\ell}}{c^2} + O(c^{-4}),
\label{eq2}
\end{equation}
where $s_{\ell}$ denotes TCL at the L-clock $\ell$.

Both gravity potentials $W_{e}$ and $W_{\ell}$ consist of the gravitational potential and the corresponding centrifugal potential. These gravity potentials include static and time-varying contributions (e.g., tidal and non-tidal effects). Following geodetic practice, we adopt the sign convention in which the leading Newtonian term of the gravitational potential is $+GM/r$; detailed definitions are given in Sec.~\ref{sec:grav-pot}.

Let $f_\ell$ denote the frequency of the electromagnetic signal measured with respect to the proper time $\tau_\ell$ of the L-clock at emission, and let $f_e$ denote the frequency of the same signal measured with respect to the proper time $\tau_e$ of the E-clock at reception. They are defined as
\[
f_\ell=\frac{1}{2\pi}\frac{d\phi}{d\tau_\ell},
\qquad
f_e=\frac{1}{2\pi}\frac{d\phi}{d\tau_e},
\]
where $\phi$ is the phase of the same electromagnetic signal, evaluated at the emission and reception events connected by the light-time equation. The conservation of phase along the light ray gives
\[
\frac{f_e}{f_\ell}
=
\frac{d\tau_\ell}{d\tau_e}.
\]
The electromagnetic signal provides a frequency-transfer observable linking the emission at the L-clock and the reception at the E-clock. The corresponding fractional frequency difference is:
\begin{equation}
\begin{aligned}
\frac{\Delta f}{f_\ell}
&=
\frac{f_e-f_\ell}{f_\ell}
=
\frac{d\tau_\ell}{d\tau_e}-1 \\
&=
\frac{d\tau_\ell}{ds_\ell}
\frac{ds_\ell}{du_\ell}
\frac{du_\ell}{du_e}
\frac{du_e}{d\tau_e}
-1 .
\end{aligned}
\label{eq:frequency_shift}
\end{equation}
The derivatives $d\tau_{\ell}/ds_{\ell}$ and $du_{e}/d\tau_{e}$ are computed from Eqs.~\eqref{eq1} and \eqref{eq2}. The term $ds_{\ell}/du_{\ell}$ is derived based on \cite{Kopeikin2024}. $du_{\ell}/du_{e}$—the derivative calculated along the electromagnetic signal path connecting the emission and reception events—is derived by solving the equations of light propagation in GCRS \cite{Kopeikin2011}. 

The fractional frequency difference between E- and L-clocks in Eq.~\eqref{eq:frequency_shift} can be rewritten as
\begin{equation}
\frac{\Delta f}{f_\ell}
=\mathcal{F}_1+\mathcal{F}_2+\mathcal{F}_3+O(c^{-4}),
\label{eq:frequency_shift_simply}
\end{equation}
where the three terms are defined as
\[
\begin{aligned}
\mathcal{F}_1 &= \frac{du_\ell}{du_e}-1,\\
\mathcal{F}_2 &= \frac{W_e-W_\ell+F}{c^2},\\
\mathcal{F}_3 &= \mathcal{F}_1\mathcal{F}_2.
\end{aligned}
\]
Here, $F/c^2$ is defined as
\[
\frac{F}{c^2}=\frac{ds_\ell}{du_\ell}-1.
\]
The terms $\mathcal{F}_1$ and $\mathcal{F}_3$ are discussed in Sec.~\ref{sec:sp}. The gravity potentials $W_e$ and $W_\ell$ in $\mathcal{F}_2$ are described in Sec.~\ref{sec:grav-pot}, while the Earth--Moon coordinate-time term $F/c^2$ is derived in Sec.~\ref{sec:coort}.

\begin{figure*}
\captionsetup{font=small}
\includegraphics[width=0.9\textwidth]{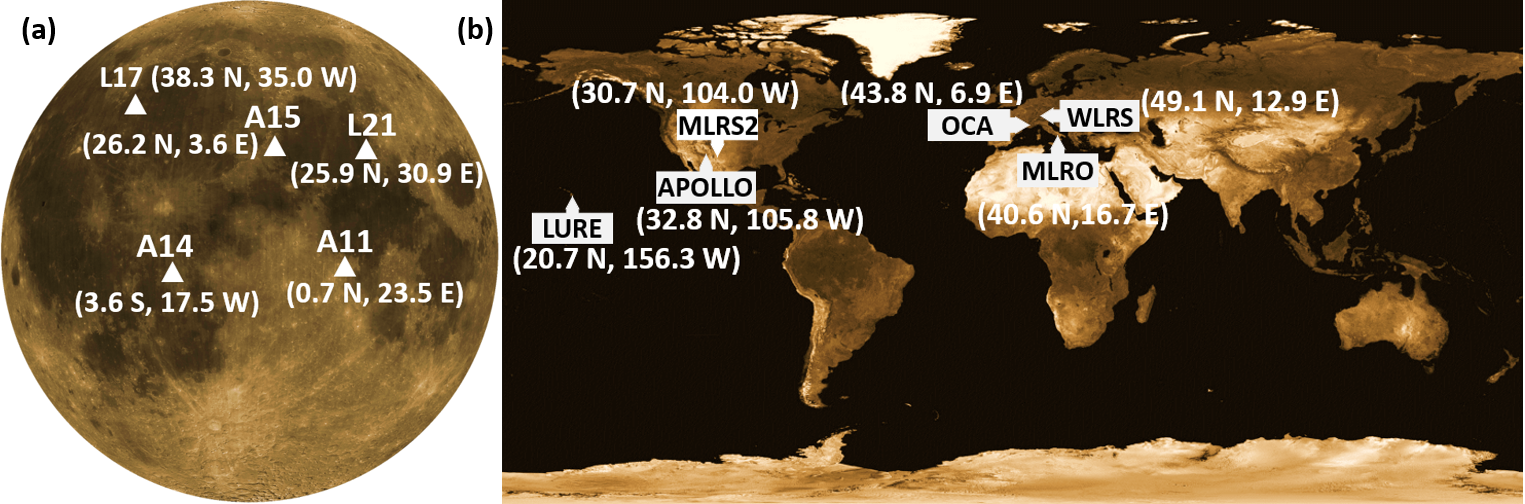}
\caption{\label{Fig1} Assumed locations of (a) L-clocks and (b) E-clocks.}
\end{figure*}

Modern optical clocks, with a fractional frequency uncertainty of $10^{-18}$, can detect gravity potential changes of $\sim$ 0.1 m$^2$/s$^2$, corresponding to $\sim$ 1 cm in height on Earth $E$ and $\sim$ 6 cm on the Moon $L$, making them vital for high-precision geodesy \cite{McGrew2018,Mueller2018}. These height changes are estimated using $\Delta W \approx g\Delta h$, with the terrestrial and lunar surface gravitational accelerations $g_{\mathrm{E}} \approx 9.81\,\mathrm{m\,s^{-2}}$ and $g_{\mathrm{L}} \approx 1.62\,\mathrm{m\,s^{-2}}$. Hourly observations are simulated by adding zero-mean Gaussian white noise with a standard deviation of $1\times10^{-18}$ to the modeled fractional frequency difference. The noise samples at different hourly epochs are statistically independent, with no temporal correlation between the hourly observations. The added noise is assumed to represent the total measurement noise of the simulated observable, rather than being treated as either optical-clock noise or frequency-transfer-link noise separately. The value of $10^{-18}$ is adopted as a representative level for this total measurement noise, with its numerical scale informed by the performance of modern optical clocks. The same numerical level of $10^{-18}$ is also adopted as a criterion for determining which physical effects are retained in the simulation. The simulation flowchart is shown in Appendix~\ref{appA}. As examples, we assume L-clocks at LLR retro-reflectors A11, A14, A15, L17, and L21, see Fig.~\ref{Fig1}(a), because their coordinates are well determined by LLR, and correspondingly, E-clocks at LLR stations LURE, APOLLO, MLRS2, OCA, WLRS, and MLRO, see Fig.~\ref{Fig1}(b). The detailed coordinates of the LLR retroreflectors and stations are given in Tab.~\ref{tab:reflector_coordinates} and Tab.~\ref{tab:station_coordinates}. The adopted LLR station coordinates and velocities and the reflector coordinates are obtained from our LLR solution using the LUNAR software \cite{Zhang2023, Singh2023, Mueller2019, Hofmann2017, Biskupek2015}. The reflector coordinates are expressed in the lunar principal axis system (PAS; \cite{Folkner2014, Park2021}). For the LLR stations, the a priori coordinates and velocities (mainly reflecting secular tectonic motion) are taken from the ITRF2014 solution \cite{Altamimi2016} for LURE, WLRS, OCA, MLRS2, and MLRO. APOLLO is not included in the ITRF2014 solution; its a priori coordinates were provided by T. Murphy \cite{Biskupek2015}. In the LLR solution, the station coordinates are determined at the reference epoch 2000.0 (2000 January 1, 00:00 UTC). The station velocities obtained from the LLR solution are used to propagate the station coordinates from the reference epoch to other epochs when required. LLR station coordinates are transformed from the terrestrial reference frame into the GCRS using the Earth orientation parameters \cite{Petit2010}, when needed, e.g., in the calculation of $\mathcal{F}_1$. LLR reflector coordinates are transformed from the PAS into the LCRS using the lunar Euler angles \cite{Folkner2014, Park2021}, and further into the GCRS or Barycentric Celestial Reference System (BCRS) using the corresponding position of the Moon, when needed, e.g., in the calculations of $\mathcal{F}_1$ and $F/c^2$.

Here, $\Delta f/f_{\ell}$ is modeled up to order $c^{-3}$. In addition, the $c^{-4}$ contribution to $F/c^{2}$ in Eq.~\eqref{eq8} is evaluated in Appendix~\ref{appC}. It lies within $[-1.0,1.4]\times10^{-18}$, i.e., at the margin of the target sensitivity. However, a consistent extension of the full $\Delta f/f_{\ell}$ model to order $c^{-4}$ would substantially increase the complexity of the formulation and is therefore left for future work. 

\begin{table*}
\caption{\label{tab:reflector_coordinates}
LLR reflector coordinates and coordinate uncertainties in the lunar principal axis system (PAS),
obtained from the LLR solution using the LUNAR software.}
\begin{ruledtabular}
\begin{tabular}{ccccccc}
Reflector
& $X$ [m] & $Y$ [m] & $Z$ [m]
& $\sigma_X$ [m] & $\sigma_Y$ [m] & $\sigma_Z$ [m] \\
\hline
A11
& 1591966.783 &  690699.387 &   21003.515
& 0.025 & 0.017 & 0.006 \\

L21
& 1339363.674 &  801871.871 &  756358.517
& 0.025 & 0.015 & 0.014 \\

A14
& 1652689.623 & -520997.634 & -109730.791
& 0.025 & 0.018 & 0.006 \\

A15
& 1554678.550 &   98095.470 &  765005.018
& 0.024 & 0.016 & 0.014 \\

L17
& 1114292.486 & -781298.476 & 1076058.529
& 0.025 & 0.013 & 0.017 \\
\end{tabular}
\end{ruledtabular}
\end{table*}

\begin{table*}
\caption{\label{tab:station_coordinates}
LLR station coordinates, coordinate uncertainties, and velocities in the terrestrial reference frame,
obtained from the LLR solution using the LUNAR software. The station coordinates refer to epoch 2000.0.}
\begin{ruledtabular}
\begin{tabular}{ccccccccccc}
Station&Site ID
& $X$ [m] & $Y$ [m] & $Z$ [m]
& $\sigma_X$ [m] & $\sigma_Y$ [m] & $\sigma_Z$ [m]
& $V_X$ [m/yr] & $V_Y$ [m/yr] & $V_Z$ [m/yr] \\
\hline
LURE&7210
& -5466007.194 & -2404427.444 & 2242188.792
& 0.007 & 0.008 & 0.015
& -0.013650 & 0.062110 & 0.032430 \\

WLRS&8834
& 4075576.815 & 931785.520 & 4801583.573
& 0.011 & 0.006 & 0.024
& -0.015640 & 0.017080 & 0.010960 \\

OCA&7845
& 4581692.145 & 556196.080 & 4389355.069
& 0.001 & 0.001 & 0.005
& -0.013607 & 0.019657 & 0.014413 \\

MLRS2&7080
& -1330021.490 & -5328403.352 & 3236481.625
& 0.004 & 0.003 & 0.008
& -0.014064 & 0.001666 & 0.003872 \\

APOLLO&7045
& -1463998.916 & -5166632.766 & 3435012.898
& 0.001 & 0.001 & 0.003
& -0.012320 & -0.000346 & -0.003269 \\

MLRO&7941
& 4641978.817 & 1393067.561 & 4133249.489
& 0.009 & 0.010 & 0.017
& -0.018810 & 0.019040 & 0.014490 \\
\end{tabular}
\end{ruledtabular}
\end{table*}

%%%%%%%%%%%%%Gravity potential%%%%%%%%%%%%%%%%%%%%%%
\section{Gravity potential}
\label{sec:grav-pot}
The gravity potential $W_q$, evaluated at a clock $q$ resting on the surface of a body $B$, consists of the gravitational potential $V_q$ and the centrifugal potential $\Phi_q$ \cite{HofmannWellenhof2006}: \[W_q=V_q+\Phi_q.\]
Here, $B\in\{E,L\}$ denotes the Earth $E$ or Moon $L$, and $q\in\{e,\ell\}$ denotes the corresponding E-clock $e$ or L-clock $\ell$. 
The gravitational potential $V_q$ consists of the contribution $V_{B,q}$ generated by body $B$ and the tidal potential $V_{\text{TP},q}$ from the external bodies \cite{Petit2010}: \[V_q=V_{B,q}+V_{\text{TP},q}.\]
 $V_{B,q}$ is further decomposed into a static part $V_{B\text{st},q}$ and a time-varying part $V_{B\text{tv},q}$: \[V_{B,q}=V_{B\text{st},q}+V_{B\text{tv},q}.\]
The static part $V_{B\text{st},q}$ is calculated using a spherical-harmonic expansion as
\begin{multline}
    V_{B\text{st},q} = \frac{GM_{B}}{r_q} 
    \Bigg[ 1 + \sum_{n=2}^\infty \left( \frac{a_{B}}{r_q} \right)^n 
    \sum_{m=0}^n \bar{P}_{nm}(\cos\theta_q) \\
    \big( \bar{C}_{nm} \cos m\lambda_q + \bar{S}_{nm} \sin m\lambda_q \big) \Bigg],
\label{eq5}
\end{multline}
where $(r_q,\theta_q,\lambda_q)$ are the spherical coordinates of clock $q$ in the body-fixed reference frame of body $B$; the fully normalized Stokes coefficients $\bar{C}_{nm}$ and $\bar{S}_{nm}$ (degree $n$ and order $m$) are from a static gravity field model of body $B$; $\bar{P}_{nm}(\cos\theta)$ is the fully normalized associated Legendre function; $G$, $M_{B}$, and $a_{B}$ represent the gravitational constant, the mass and equatorial radius of body $B$ \cite{HofmannWellenhof2006}. In the numerical calculations, the spherical-harmonic expansion in Eq.~\eqref{eq5} is truncated at degree 2190 for the terrestrial EIGEN-6C4 model and at degree 500 for the lunar GL0900D model, as detailed in Secs.~\ref{sec:grav-pot-E} and~\ref{sec:grav-pot-L}.

We group the time-varying part $V_{B\text{tv},q}$, including the tidal and non-tidal contributions, and the external tidal potential $V_{\text{TP},q}$ into effective potential variations (EPVs), defined by \[\sum V_{\text{EPV},q}=V_{B\text{tv},q}+V_{\text{TP},q},\] where the sum is taken over the individual contributions. An EPV due to a forcing can be expanded in spherical harmonics as
\begin{equation}
V_{\text{EPV},q}= \sum_{n} \bigl(1+k_n-h_n\bigr)\, V^{\rm dir}_{n}. 
\end{equation}
Here $V^{\rm dir}_{n}$ is the degree-$n$ direct potential of the forcing, e.g., the tidal potential $V_{\text{TP},q}$, and the load potential from the mass redistribution of body $B$; $\left(k_n-h_n\right)V^{\rm dir}_{n}$ is the response of body $B$ to this forcing, combining the induced self-potential change ($k_n$ term) and surface deformations ($h_n$ term) \cite{Voigt2016,Schroeder2021,Vincent2024a}. $k$ and $h$ are Love numbers; or load Love numbers for the loading potentials \cite{Farrell1972,Love1911}.

The centrifugal potential $\Phi_{q}$ is determined by the angular velocity $\bm{\omega}_{B}$ of body $B$ and the position $\bm{x}_q$ of clock $q$, here expressed in the body-fixed reference frame of $B$:
\begin{equation}
\Phi_{q} = \frac{1}{2} (\bm{\omega}_{B} \times \bm{x}_q)^2. 
\label{c}
\end{equation} 
%%%%%%%%%%%%%Gravity potential: E-clocks%%%%%%%%%%%%%%%%%%%
\subsection{Gravity potential magnitudes for E-clocks}
\label{sec:grav-pot-E}
For E-clocks ($q=e$) on the Earth ($B=E$), the EIGEN-6C4 model \cite{Foerste2014} is adopted to compute the static gravitational potential $V_{E\text{st},e}$ generated by the Earth. Using all normalized Stokes coefficients to degree 2190, EIGEN-6C4 provides $V_{E\text{st},e}/c^2$ uncertainties similar to or lower than those of the alternative models and is the only one keeping the uncertainties at all six E-clocks on the $10^{-18}$ level (Tab.~\ref{Tab4}). Truncating to degree 200 further reduces the $V_{E\text{st},e}/c^2$ uncertainties to below $1\times10^{-18}$. However the contribution to $V_{E\text{st},e}/c^2$ from degrees 201--2190 cannot be omitted because its value can be on the $10^{-17}$ or even $10^{-16}$ level at some sites and is statistically significant relative to its uncertainty. We therefore retain the full field. For sites targeting $10^{-18}$ accuracy, global models alone are close but still insufficient; adding site-specific/local data is a practical path \cite{Delva2019,Grotti2024}. Inter-model differences in $V_{E\text{st},e}/c^2$ are on or below the $10^{-18}$ level (Tab.~\ref{Tab5}). 

Effective potential variation (EPV) of tidal factors $V_{\text{EPVT},e}$ includes the effects of the solid Earth tide and pole tide, computed using ETERNA 3.4 \cite{Wenzel1996}, with the factor $(1+k-h)$ applied to obtain the corresponding EPV, and the effect of the ocean tide loading from SPOTL 3.3.0.2 \cite{Agnew2012, Agnew1997}. Other tidal factors, such as ocean pole tide loading and atmospheric tide loading, are below the adopted threshold for the sites and models considered. Their contributions to the fractional frequency shift are approximately $[-1.5,\,1.6]\times 10^{-19}$ and $[-2.3,\,2.4]\times 10^{-19}$. For the non-tidal EPV, $V_{\text{EPVNT},\,e}$, (i) the contribution of $(1+k)$ term is quantified by the daily ITSG-Grace2018 models \cite{MayerGuerr2018, Kvas2019}, retaining only the time-variable hydrological, cryospheric, and glacial isostatic adjustment signals after subtracting the static potential part, and by the daily AOD1B RL06 time series \cite{Shihora2022}, containing the non-tidal atmospheric and oceanic effects; (ii) the $h$ term is calculated based on daily GNSS vertical displacement series \cite{Blewitt2018} of selected stations MAUI, WTZR, GRAS \cite{RESIF2017}, MDO1, P027 \cite{Community2007} and MATE, including the non-tidal effects from hydrology, atmosphere, ocean, etc. 

Earth's angular velocity used for calculating the centrifugal potential $\Phi_{e}$ is derived from IERS EOP C04 LOD series \cite{Bizouard2019}.

Each component of the E-clock gravity potential $W_{e}$—$V_{E\text{st},e}$, $V_{\text{EPVT},e}$, $V_{\text{EPVNT},e}$ and $\Phi_{e}$—contributes to $\Delta f/f_{\ell}$ by its value divided by $c^2$. To obtain more accurate and complete variation ranges, $\Phi_{e}/c^2$ is calculated using the IERS EOP C04 series from 1962 to 2025; the tidal and non-tidal EPVs are evaluated over 2002-2017 to match the full ITSG-Grace2018 model span. 

The dominant terms are $V_{E\text{st},e}/c^2$ (about $6.96 \times 10^{-10}$) and $\Phi_{e}/c^2$ ($[5.1, 10.5]\times10^{-13}$) for the six E-clocks (Tab.~\ref{Tab1}). Tidal contribution is $[-2.8,\,4.8]\times10^{-17}$, mainly from the EPV of solid Earth tide ($[-2.4,\,4.6]\times 10^{-17}$) while the EPVs of the ocean tide loading and solid Earth pole tide contribute $[-8.5,\,10.3]\times 10^{-18}$ and $[-8.8,\,9.0]\times 10^{-19}$. Non-tidal contribution based on the AOD, ITSG and GNSS series varies within $[-2.9,\,3.3]\times 10^{-18}$ for the E-clocks, in which the seasonal signal fitted ranges within $[-6.3,\,5.1]\times 10^{-19}$, mainly due to the hydrology, atmosphere and ocean (with similar level obtained by HYDL, NTAL and NTOL loading models from GFZ \cite{Dill2013}).

\begin{table}
\caption{\label{Tab1}Effects of $W_e$ components on $\Delta f / f_\ell$.}
\begin{ruledtabular}
\begin{tabular}{cccc} 
\shortstack{$V_{E\text{st},e}/c^2$\\$\times 10^{-10}$} &
\shortstack{$\Phi_{e}/c^2$\\$\times 10^{-13}$} &
\shortstack{$V_{\text{EPVT},e}/c^2$\\$\times 10^{-17}$} &
\shortstack{$V_{\text{EPVNT},e}/c^2$\\$\times 10^{-18}$}\\
\midrule
7.0& [5.1,\,10.5]& [-2.8,\,4.8]&[-2.9,\,3.3] \\
\end{tabular}
\end{ruledtabular}
\end{table}
%%%%%%%%%%%%%Gravity potential: L-clocks%%%%%%%%%%%%%%%%%%%
\subsection{Gravity potential magnitudes for L-clocks}
\label{sec:grav-pot-L}
For L-clocks ($q=\ell$) on the Moon ($B=L$), the GL0900D model \cite{Konopliv2014,Kahan2022} is used for calculating the static gravitational potential $V_{L\text{st},\ell}$ generated by the Moon. GL0900D is preferred among ten models: its coefficients can be retained up to a threshold degree $n_{thr}=500$ for a more reliable result; using degrees $n\le500$ yields the most accurate $V_{L\text{st},\ell}/c^2$ with uncertainty at $10^{-19}$ level (Tab.~\ref{Tab6}). Degrees $n>n_{thr}$ are noise-dominated (Fig.~\ref{Fig5}) and thus excluded; their contribution to $V_{L\text{st},\ell}/c^2$ is $>1\times10^{-18}$ but insignificant compared to 1- or 2-fold of its uncertainty, i.e., A11 $(0.6\pm1.4)\times10^{-17}$, L21 $(0.5\pm2.2)\times10^{-17}$, A14 $(-0.7\pm1.0)\times10^{-17}$, A15 $(2.7\pm1.4)\times10^{-17}$ and L17 $(2.5\pm1.7)\times10^{-17}$. Nine other models behave similarly. This reflects that the model-limited systematic noise floor of lunar global models alone at a given site may remain above $10^{-18}$ without additional local/site-specific gravity information, as in the Earth case. Including noisy coefficients inflates inter-model differences of $V_{L\text{st},\ell}/c^2$, while excluding them keeps differences on or below the $10^{-18}$ level (Tab.~\ref{Tab7}). More details are in Appendix~\ref{appB}. 

The tidal potential $V_{\text{TP},\ell}$ at the L-clock $\ell$ with spherical coordinates $(r_{\ell}, \theta_{\ell}, \lambda_{\ell})$ resulting from a tide-generating body $b$ with mass $M_b$ and coordinates $(r_b, \theta_b, \lambda_b)$ at time $t$ can be calculated by \cite{Wenzel1997}
\begin{multline}
    V_{\text{TP},\ell} = \frac{GM_b}{r_b} \sum_{n=2}^{\infty} \left( \frac{r_{\ell}}{r_b} \right)^n \frac{1}{2n + 1} \sum_{m=0}^{n} \bar{P}_{nm}(\cos \theta_{\ell}) \\
  \bar{P}_{nm}(\cos \theta_b)\, \cos\big( m\lambda_{\ell} - m\lambda_b \big).
\label{eq7}
\end{multline}
Here, both sets of spherical coordinates are expressed in the lunar body-fixed reference frame. The bodies' ephemerides are generated by numerically integrating a DE440-based dynamical model \cite{Park2021} in the LLR analysis software LUNAR \cite{Zhang2023, Singh2023, Mueller2019, Hofmann2017, Biskupek2015}. Given the clock uncertainty of $10^{-18}$, only considering Earth and Sun contributions up to degrees $n=3$ and 2 is sufficient; higher degrees and other bodies can be negligible. For the effective potential variation $V_{\text{EPVT},\ell}$, lunar degree-2 Love numbers ($k_{L2}$ and $h_{L2}$) and the associated tidal time delay are considered.

The angular velocity of the lunar mantle $\bm{\omega}_L$ is obtained by integrating its time derivative \cite[Eq.~(34)]{Folkner2014} in the LUNAR software and is used to calculate the lunar centrifugal potential $\Phi_{\ell}$.

Tab.~\ref{Tab2} summarizes the contributions of $W_{\ell}$ elements to $\Delta f/f_{\ell}$: $V_{L\text{st},\ell}/c^2$ about $3.14 \times 10^{-11}$, $\Phi_{\ell}/c^2$ around $1 \times 10^{-16}$, and $V_{\rm EPVT,\ell}/c^2$ within $[-0.3, 2.6]\times10^{-16}$ for the five L-clocks. Variation ranges of $V_{\rm EPVT,\ell}/c^2$ and $\Phi_{\ell}/c^2$ are derived from over 50 years of ephemerides and a $\bm{\omega}_L$ series. In $V_{\rm EPVT,\ell}/c^2$, the dominant contribution comes from Earth (up to degree 3), while effects from the Sun (degree 2) and lunar response ($k_{L2}$ and $h_{L2}$ part) are smaller, within $[-0.7, 1.4] \times 10^{-18}$ and $[-5.0, 0.6] \times 10^{-18}$. Earth degree 3 alone contributes $[-0.6, 1.2] \times 10^{-18}$. 

\begin{table}
\captionsetup{font=small}
\caption{\label{Tab2}Effects of $W_\ell$ components on $\Delta f/f_\ell$.}
\begin{ruledtabular}
\begin{tabular}{ccc} 
\shortstack{$V_{L\rm st,\ell}/c^2$\\$\times 10^{-11}$} &
\shortstack{$\Phi_{\ell}/c^2$\\$\times 10^{-16}$} &
\shortstack{$V_{\rm EPVT,\ell}/c^2$\\$\times 10^{-16}$} \\
\midrule
3.1 & [0.7,\,1.2] & [-0.3,\,2.6] \\
\end{tabular}
\end{ruledtabular}
\end{table}
%%%%%%%%%%%%%Earth-Moon coordinate time ratio%%%%%%%%%%%%%%%%%%%
\section{Earth--Moon coordinate time ratio}
\label{sec:coort}
The relationships between Barycentric Coordinate Time ($t\equiv$ TCB) and Geocentric Coordinate Time ($u\equiv$ TCG), and between TCB and Lunar Coordinate Time ($s\equiv$ TCL) follow \cite[Eqs. (1) and (2)]{Kopeikin2024}.  At the L-clock $\ell$, $F/c^2=(ds_\ell/dt)(dt/du_\ell)-1$. Computing the derivatives, we obtain
\begin{multline}
    \frac{F}{c^2} = 
    \frac{1}{c^2} \Bigl(\sum_{A \neq E} \frac{GM_A}{r_{EA}}-\sum_{A \neq L} \frac{GM_A}{r_{LA}}+\frac{v_E^2 - v_L^2}{2}+\bm{a}_E \cdot \bm{r}_{\ell E}  \\   
    +\bm{v}_E \cdot \bm{v}_{\ell E} -\bm{a}_L \cdot \bm{r}_{\ell L} - \bm{v}_L \cdot \bm{v}_{\ell L}\Bigr)+O(c^{-4}). 
\label{eq8}
\end{multline}
We use BCRS coordinates ${\bm x}_E$, ${\bm x}_L$, ${\bm x}_\ell$ of the Earth $E$, Moon $L$ and L-clock $\ell$, and their velocities ${\bm v}_E$, ${\bm v}_L$, and ${\bm v}_\ell$. 
We denote  
\[
\begin{aligned}
\bm{r}_{\ell E} &= \bm{x}_\ell-\bm{x}_E,
&\qquad \bm{r}_{\ell L} &= \bm{x}_\ell-\bm{x}_L,\\
\bm{v}_{\ell E} &= \bm{v}_\ell-\bm{v}_E,
&\qquad \bm{v}_{\ell L} &= \bm{v}_\ell-\bm{v}_L.
\end{aligned}
\]
The BCRS accelerations of the Earth and Moon are $\bm{a}_E$ and $\bm{a}_L$. The distances $r_{EA}$ and $r_{LA}$ are between body $A$ (with mass $M_A$) and Earth, and between body $A$ and Moon. The effects from the body position uncertainty and numerical integration error are below $10^{-18}$, see Appendix~\ref{appC}. 
 
As an independent consistency check of Eq.~\eqref{eq8}, in which $F/c^2$ is obtained through TCB as an intermediate timescale, we also derive $F/c^2$ by differentiating the direct TCG--TCL relation in \cite[Eq.~(12)]{Kopeikin2024}, where most BCRS coordinate-related terms cancel.    
\begin{multline}
\frac{F}{c^2} = -\frac{1}{c^2}\Bigl(\bm{a}_{LE} \cdot \bm{r}_{\ell L} + \bm{v}_{LE} \cdot \bm{v}_{\ell L}+\frac{v_{LE}^2}{2} \\
+ \frac{G M_E}{r_{LE}} - \frac{2G M_L}{r_{LE}}+U_2+U_3+\sum_{j>3} U_j\Bigr)+O(c^{-4}),
\label{eq9}
\end{multline} where \[\begin{aligned}\bm{r}_{LE} &= \bm{x}_L-\bm{x}_E, &\ \bm{v}_{LE} &= \bm{v}_L-\bm{v}_E,&\ \bm{a}_{LE} &= \bm{a}_L-\bm{a}_E.\end{aligned}\] The tidal-potential effects of the second-order $U_2$ and third-order $U_3$ are
\begin{equation}
U_2=\sum_{A\neq E,L}\frac{GM_A}{2r_{EA}}
\left(\frac{r_{LE}}{r_{EA}}\right)^2
\left(3\cos^2\psi_{LA}-1\right),
\label{eq10}
\end{equation}
and
\begin{equation}
U_3=\sum_{A\neq E,L}\frac{GM_A}{2r_{EA}}
\left(\frac{r_{LE}}{r_{EA}}\right)^3
\left(5\cos^3\psi_{LA}-3\cos\psi_{LA}\right).
\label{eq11}
\end{equation}
Here, $\cos\psi_{LA}={\bm n}_{AE} \cdot {\bm n}_{LE}$, where $\psi_{LA}$ is the angle between the unit vectors ${\bm n}_{AE}$ and ${\bm n}_{LE}$, directed from Earth's geocenter to body $A$ and to the Moon. The higher-order tidal terms $\sum_{\scriptscriptstyle {j>3}}U_j$ give contributions to $F/c^2$ below the $10^{-18}$ level and can therefore be neglected here in the evaluation. 

Compared with \cite[Eq.~(12) and its variant Eq.~(29)]{Kopeikin2024}, our Eq.~\eqref{eq9} further considers $U_3$. This term is relevant here because $U_3/c^2$ ranges within $[-2.1, 2.1]\times10^{-16}$ for the assumed L-clocks. The analytic solution of \cite[Eq.~(29)]{Kopeikin2024}, given in \cite[Eq.~(72)]{Kopeikin2024}, relies on a simplified Keplerian model and neglects perturbations from the major planets, whereas our approach uses numerically integrated ephemerides to capture the full multi-body dynamics. Kopeikin and Kaplan \cite[Eq.~(99)]{Kopeikin2024} explicitly expressed how lunar geometric libration affects the lunar--terrestrial time transformation. In our formulations, this effect is implicitly included via ephemerides and the light-time solution. Physical libration is included through the lunar-orientation model used to transform coordinates from the lunar body-fixed frame to the BCRS. Thus, Eqs.~\eqref{eq8} and \eqref{eq9} incorporate the relevant libration effects via the L-clock positions and Earth–Moon distance.

Barycentric Dynamical Time (TDB) is used in LUNAR for calculating the ephemerides, while the quantities in Eqs.~\eqref{eq8} and \eqref{eq9} are TCB-compatible. TDB is scaled to TCB by the factor $(1-L_B)$, where $L_B=1.550519768 \times 10^{-8}$ \cite{Petit2010}, along with corresponding re-scaling of spatial coordinates and body masses. This procedure leaves Eqs.~\eqref{eq8} and \eqref{eq9} unchanged \cite{Klioner2008}. 

\begin{figure}
\captionsetup{font=small}
\includegraphics[width=\columnwidth]{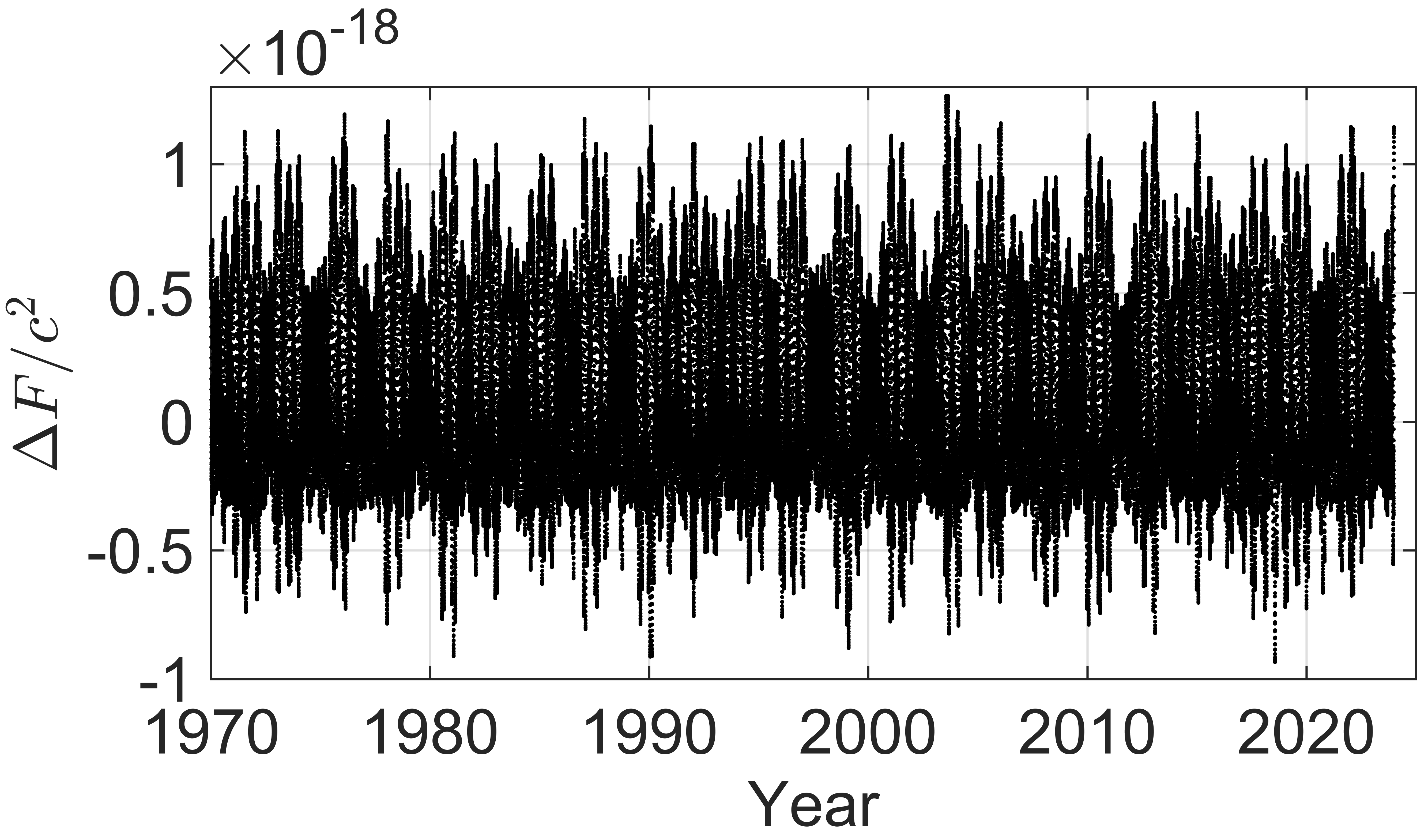}
\caption{\label{Fig2}Dimensionless difference $\Delta F/c^2 =(F/c^2)_{Eq.\eqref{eq8}}-(F/c^2)_{Eq.\eqref{eq9}}$ over 50 years, with the L-clock assumed to be located at A15. Eq.~\eqref{eq8} includes the Sun, Moon, and eight planets. Eq.~\eqref{eq9} includes the Sun, Moon, Earth, Venus, and Jupiter.}
\end{figure}

Both Eqs.~\eqref{eq8} and \eqref{eq9} yield $F/c^2$ in the range $[-1.9, -1.6]\times10^{-11}$. For the five L-clock sites, the mean values obtained from Eq.~\eqref{eq8} are all approximately $-1.70938\times10^{-11}$, while the site-dependent ranges are listed in Tab.~\ref{tab:F_c2}. For the $10^{-18}$ clock uncertainty, Eq.~\eqref{eq8} requires all ten major solar-system bodies (Moon, Sun, and eight planets), while the body contributions listed in Tab.~\ref{Tab9} indicate that Eq.~\eqref{eq9} requires the inclusion of the Earth, Moon, Sun, Venus, and Jupiter. As shown in Fig.~\ref{Fig2}, the differences between Eqs.~\eqref{eq8} and \eqref{eq9} lie within $[-1,\,1]\times10^{-18}$ at nearly all epochs over more than 50 years. Although Eqs.~\eqref{eq8} and \eqref{eq9} are derived through different formalisms; they describe the same coordinate-time transformation effect. The agreement between their numerical evaluations at the $10^{-18}$ level demonstrates the internal consistency of the relativistic Earth--Moon reference-system framework adopted in this work. Both equations indicate $10^{-15}$-level $F/c^2$ differences among L-clock sites. In both equations, Earth, Moon, and Sun dominate, followed by Jupiter and Venus; the other bodies contribute less. Per-body contributions (Tab.~\ref{Tab9}) differ between the two equations because body-specific terms are defined differently. In contrast to Eq.~\eqref{eq8}, in Eq.~\eqref{eq9}, the contributions are: Venus $>$ Jupiter and Mars $>$ Saturn, as the distance dominates mass in tidal potentials. 

\begin{table}[t]
\caption{\label{tab:F_c2}
Ranges of $F/c^2$ obtained from Eq.~\eqref{eq8} for the five L-clock sites over more than 50 years.}
\begin{ruledtabular}
\begin{tabular}{cc}
L-clock site & $F/c^2~(\times10^{-11})$ \\
\hline
A11 & $[-1.89766,\,-1.58434]$ \\
A14 & $[-1.89768,\,-1.58430]$ \\
A15 & $[-1.89788,\,-1.58425]$ \\
L17 & $[-1.89792,\,-1.58422]$ \\
L21 & $[-1.89782,\,-1.58427]$ \\
\end{tabular}
\end{ruledtabular}
\end{table}

\begin{table}
\captionsetup{font=small}
\caption{\label{Tab9}Per-body contributions to $F/c^2$ in Eqs.~\eqref{eq8} and \eqref{eq9}. The per-body entries in the second and third columns are not directly comparable term by term because $F/c^2$ is decomposed differently in the two equations.}    
\begin{ruledtabular}
\begin{tabular}{ccc}
\shortstack[c]{Body\\ {}}&\shortstack[c]{Contribution to $F/c^2$\\ in Eq.~\eqref{eq8}} & \shortstack[c]{Contribution to $F/c^2$ \\in Eq.~\eqref{eq9}} \\
\hline
 Earth+Moon& $[-4.4,\,1.2]\times10^{-11}$ &  $[-1.9,\,1.6]\times10^{-11}$\\  
 Sun& $[-2.8,\, 2.8]\times10^{-11}$ &$[-7.7,\,3.8]\times10^{-14}$\\
    Jupiter & $[-1.6,\,1.6]\times10^{-15}$ & $[-1.1,\,0.6]\times10^{-18}$ \\
    Venus   & $[-9.2,\,9.0]\times10^{-16}$ & $[-9.2,\,4.6]\times10^{-18}$ \\
    Saturn  & $[-1.2,\,1.2]\times10^{-16}$ & $[-4.0,\,2.0]\times10^{-20}$ \\
    Mars    & $[-5.8,\,5.8]\times10^{-17}$ & $[-4.0,\,2.2]\times10^{-19}$ \\
    Mercury & $[-1.5,\,1.4]\times10^{-17}$ & $[-7.2,\,3.5]\times10^{-20}$ \\
    Uranus  & $[-3.9,\,3.8]\times10^{-18}$ & $[-6.0,\,3.0]\times10^{-22}$ \\
    Neptune & $[-1.6,\,1.6]\times10^{-18}$ & $[-1.5,\,0.8]\times10^{-22}$ \\
\end{tabular}
\end{ruledtabular}
\end{table}
%%%%%%%%%%%%%Signal propagation%%%%%%%%%%%%%%%%%%%
\section{Signal propagation}
\label{sec:sp}
In the present simulations, the E-L clock link is assumed to be an optical laser link. We derive $\mathcal{F}_1=du_\ell/du_e-1$ from the one-way ranging light-time equation for the L-clock-to-E-clock signal in the GCRS, corresponding to the Moon-to-Earth leg of the LLR two-way light-time model \cite{Zhang2023, Singh2023, Mueller2019, Hofmann2017, Biskupek2015}. Here $\mathcal{F}_1$ includes the contributions from the Doppler shift $\mathcal{F}_{1\rm dp}$, Shapiro delay $\mathcal{F}_{1\rm sha}$, and tropospheric delay $\mathcal{F}_{1\rm trop}$. Ionospheric delay is neglected because it is negligible for the optical laser link considered here. Additional foundational treatments of relativistic time and frequency transfer are given in \cite {Blanchet2001} and \cite{ Linet2002}. 

The Doppler terms are derived as
\begin{multline}
\mathcal{F}_{1\rm dp}=-\frac{ \bm{n}_{e \ell}\cdot\bm{v}_{e \ell}}{c} - \frac{(\bm{n}_{e \ell}\cdot\bm{v}_{e \ell})(\bm{n}_{e \ell}\cdot\bm{v}_{\ell})}{c^2}+O(c^{-3}),
\label{eqD1}
\end{multline}
where the unit vector $\bm{n}_{e \ell}$ is directed from the L-clock $\ell$ to the E-clock $e$; $\bm{v}_{e \ell}=\bm{v}_{e}-\bm{v}_{\ell}$, with $\bm{v}_{e}$ and $\bm{v}_{\ell}$ the geocentric velocities of $e$ and $\ell$. $c^{-3}$ and higher orders can be ignored for the $10^{-18}$ level.

From \cite[Eq.~(11.17)]{Petit2010}, we derive Shapiro terms in $\mathcal{F}_1$ for Earth ($E$):
\begin{multline}
\mathcal{F}_{1\rm sha}= \frac{\alpha_E}{r_1r_2}[(r_1-r_2)(\bm{n}_{e} \cdot \bm{v}_e + \bm{n}_{\ell} \cdot \bm{v}_{\ell})-(r_1+r_2)\bm{n}_{e \ell} \cdot \bm{v}_{e \ell}]\\
+O(c^{-4}),
\label{eqD2}
\end{multline}
where $\alpha_E=2GM_E/c^3$; the unit vectors $\bm{n}_{e}$ and $\bm{n}_{\ell}$ point from geocenter to the E-clock $e$ and the L-clock $\ell$. The quantities $r_1$ and $r_2$ are defined as \[r_1=r_e+r_\ell+r_{e\ell},\qquad r_2=r_e+r_\ell-r_{e\ell},\]where $r_{e}$ and $r_{\ell}$ are the geocentric distances of $e$ and $\ell$, and $r_{e \ell}$ is the distance between $e$ and $\ell$. 
At the $10^{-18}$ level, retaining $c^{-3}$ terms suffices. In the GCRS, other bodies contribute through tidal effects. Their contributions lie only within $[-3.2, 3.2]\times10^{-19}$ for all assumed clocks and are therefore neglected here.

Based on \cite[section 9.1]{Petit2010}, the tropospheric terms are derived as
\begin{multline}
\mathcal{F}_{1\rm trop}= \frac{d_z\,m'(\epsilon)}{c\,\cos\epsilon}\left(\dot{\bm{n}}_{U} \cdot \bm {n}_{e \ell}+\bm{n}_{U} \cdot \frac{(\bm{I}-\bm{n}_{e \ell}\bm{n}_{e \ell}^{T})\bm{v}_{e \ell}}{r_{e \ell}}\right)\\-\frac{m(\epsilon)\dot{d}_z}{c}+O(c^{-2}),
\label{eqD3}
\end{multline}
where $\dot{d}_z$ and $\dot{\bm{n}}_{U}$ are the time derivatives of the zenith delay $d_{z}$ \cite[Eqs.~(9.2)-(9.7)]{Petit2010} and of the zenith unit vector $\bm{n}_{U}$ at the E-clock. $m'(\epsilon)=dm(\epsilon)/d\epsilon$, where $m(\epsilon)$ is the FCULa mapping function developed by Mendes et al.~\cite{Mendes2002} and adopted in the IERS Conventions 2010~\cite{Petit2010}. The FCULa mapping function is valid for elevation angles $\epsilon>3^\circ$; therefore, this lower elevation limit is adopted in the present simulations. The low-elevation performance of the FCULa mapping function is assessed in \cite{Mendes2002}. The meteorological inputs used to compute $d_z$ and $\dot{d}_z$ for the six E-clock sites are surface pressure, temperature, and relative humidity from the ERA5-Land dataset \cite{MunozSabater2019,MunozSabater2021}, accessed through the Open-Meteo Historical Weather API \cite{Zippenfenig2023}. Since $\mathbf{n}_{e\ell}$ is directed from the L-clock to the E-clock, whereas the line of sight defining the elevation angle is directed from the E-clock to the L-clock, $\sin\epsilon=-\mathbf{n}_U\cdot\mathbf{n}_{e\ell}$. Therefore, Eq.~\eqref{eqD3} is consistent with $\mathcal{F}_{1\rm trop}=-d[d_zm(\epsilon)/c]/du_e$. Terms of order $c^{-2}$ and higher can be ignored here.

The magnitude of $\mathcal{F}_{1\rm trop}$ decreases with increasing elevation angle $\epsilon$. Over more than 50 years, for all assumed E- and L-clocks, its overall range is $[-7.5,\,7.5]\times 10^{-11}$ for $3^\circ < \epsilon < 30^\circ$ and $[-1.4,\,1.4]\times 10^{-12}$ for $\epsilon \ge 30^\circ$. The magnitudes of both $\mathcal{F}_{1\rm dp}$ and $\mathcal{F}_{1\rm sha}$, at the $10^{-6}$ and $10^{-15}$ levels, respectively, reduce with increasing E-clock latitude and elevation angle $\epsilon$. This behavior is illustrated in Tab.~\ref{tab:F1_components} for all assumed E-clock sites. For all clocks, the combined signal-propagation term $\mathcal{F}_1=\mathcal{F}_{1\rm dp}+\mathcal{F}_{1\rm sha}+\mathcal{F}_{1\rm trop}$ ranges within $[-1.7,\,1.7]\times 10^{-6}$ for $\epsilon>3^\circ$ and the cross-term $\mathcal{F}_3=\mathcal{F}_1\mathcal{F}_2$, combining the effects from the signal propagation, gravity potential and coordinate time ratio, varies within $[-1.1,\,1.1]\times 10^{-15}$. 

\begin{table*}[t]
\caption{\label{tab:F1_components}
Ranges of $\mathcal{F}_{1\rm dp}$, $\mathcal{F}_{1\rm sha}$, and $\mathcal{F}_{1\rm trop}$ for the six E-clock sites
evaluated over a period of more than 50 years, separated into the
elevation-angle intervals $3^\circ<\epsilon<30^\circ$ and
$\epsilon\geq30^\circ$.}
\begin{ruledtabular}
\begin{tabular}{lcccccc}
& \multicolumn{2}{c}{$\mathcal{F}_{1\rm dp}$ (all entries $\times10^{-6}$)}
& \multicolumn{2}{c}{$\mathcal{F}_{1\rm sha}$ (all entries $\times10^{-15}$)}
& \multicolumn{2}{c}{$\mathcal{F}_{1\rm trop}$ (all entries $\times10^{-12}$)}
\\
Station (latitude)
& $3^\circ<\epsilon<30^\circ$
& $\epsilon\geq30^\circ$
& $3^\circ<\epsilon<30^\circ$
& $\epsilon\geq30^\circ$
& $3^\circ<\epsilon<30^\circ$
& $\epsilon\geq30^\circ$
\\
\hline
LURE ($20.7^\circ$ N)
& $[-1.65,\,1.65]$
& $[-1.45,\,1.45]$
& $[-1.87,\,1.87]$
& $[-1.15,\,1.15]$
& $[-69.46,\,69.51]$
& $[-1.32,\,1.31]$
\\
MLRS2 ($30.7^\circ$ N)
& $[-1.54,\,1.54]$
& $[-1.35,\,1.35]$
& $[-1.73,\,1.73]$
& $[-1.06,\,1.06]$
& $[-71.48,\,71.34]$
& $[-1.37,\,1.36]$
\\
APOLLO ($32.8^\circ$ N)
& $[-1.52,\,1.51]$
& $[-1.34,\,1.34]$
& $[-1.69,\,1.69]$
& $[-1.04,\,1.03]$
& $[-64.25,\,64.43]$
& $[-1.21,\,1.21]$
\\
MLRO ($40.6^\circ$ N)
& $[-1.39,\,1.40]$
& $[-1.21,\,1.22]$
& $[-1.53,\,1.53]$
& $[-0.94,\,0.93]$
& $[-75.00,\,74.77]$
& $[-1.44,\,1.43]$
\\
OCA ($43.8^\circ$ N)
& $[-1.34,\,1.34]$
& $[-1.18,\,1.17]$
& $[-1.46,\,1.46]$
& $[-0.89,\,0.89]$
& $[-65.26,\,65.37]$
& $[-1.24,\,1.24]$
\\
WLRS ($49.1^\circ$ N)
& $[-1.24,\,1.24]$
& $[-1.08,\,1.09]$
& $[-1.33,\,1.33]$
& $[-0.81,\,0.80]$
& $[-64.37,\,64.39]$
& $[-1.23,\,1.22]$
\\
\end{tabular}
\end{ruledtabular}
\end{table*}

For a single link, higher elevations and higher E-clock latitudes reduce signal-propagation effects. However, using multiple links is more practical. The triple-link Doppler-cancelling scheme \cite{Vessot1979,Shen2016,Shen2023} combines a coherent two-way transponded link with an independent one-way clock link. Here, we consider two configurations: Earth–Moon–Earth and Moon–Earth–Moon.

For the Earth--Moon--Earth configuration, the E-clock transmits a signal at $t_1$, which is received by the L-clock at $t_{2,r}$. The received signal is coherently retransmitted toward the Earth at $t_{2,c}$, while an independent signal generated by the L-clock is emitted at $t_{2,i}$. The coherent and independent downlink signals are received by the E-clock at $t_{3,c}$ and $t_{3,i}$, respectively. The corresponding frequencies are denoted by $f_1$, $f_{2,r}$, $f_{2,c}$, $f_{2,i}$, $f_{3,c}$, and $f_{3,i}$.

Following the four time transformations in Eq.~\eqref{eq:frequency_shift}, the fractional-frequency shifts for the uplink (Earth to Moon), coherent downlink (Moon to Earth), and independent downlink (Moon to Earth) are
\begin{subequations}
\label{eq:eme_three_links}
\begin{equation}
\label{eq:eme_uplink}
\begin{aligned}
\frac{f_{2,r}}{f_1}-1={}&\left(\frac{d\tau_e}{du_e}\right)_1\frac{du_{e,1}}{du_{\ell,2r}}\left(\frac{du_\ell}{ds_\ell}\right)_{2r}\\
&\times\left(\frac{ds_\ell}{d\tau_\ell}\right)_{2r}-1,
\end{aligned}
\end{equation}
\begin{equation}
\label{eq:eme_coherent_downlink}
\begin{aligned}
\frac{f_{3,c}}{f_{2,c}}-1={}&\left(\frac{d\tau_\ell}{ds_\ell}\right)_{2c}\left(\frac{ds_\ell}{du_\ell}\right)_{2c}\\
&\times\frac{du_{\ell,2c}}{du_{e,3c}}\left(\frac{du_e}{d\tau_e}\right)_{3c}-1,
\end{aligned}
\end{equation}
\begin{equation}
\label{eq:eme_independent_downlink}
\begin{aligned}
\frac{f_{3,i}}{f_{2,i}}-1={}&\left(\frac{d\tau_\ell}{ds_\ell}\right)_{2i}\left(\frac{ds_\ell}{du_\ell}\right)_{2i}\\
&\times\frac{du_{\ell,2i}}{du_{e,3i}}\left(\frac{du_e}{d\tau_e}\right)_{3i}-1.
\end{aligned}
\end{equation}
\end{subequations}

The triple-link fractional-frequency observable is formed by subtracting one half of the coherent two-way fractional-frequency shift from the independent one-way fractional-frequency shift:
\begin{equation}
\label{eq:triple_observable}
\begin{aligned}
y_{\mathrm{tri}}=\left(\frac{f_{3,i}}{f_{2,i}}-1\right)-\frac12\left(\frac{f_{3,c}}{kf_1}-1\right),
\end{aligned}
\end{equation}
where $k=f_{2,c}/f_{2,r}$ denotes the effective frequency turnaround ratio of the coherent transponder.

Substituting the three frequency-transfer ratios in Eqs.~\eqref{eq:eme_uplink}--\eqref{eq:eme_independent_downlink} into the triple-link observable in Eq.~\eqref{eq:triple_observable} gives the general expression for the Earth--Moon--Earth configuration:
\begin{equation}
\label{eq:eme_general}
\begin{aligned}
y_{\mathrm{tri}}={}&
\left(\frac{d\tau_\ell}{ds_\ell}\right)_{2i}
\left(\frac{ds_\ell}{du_\ell}\right)_{2i}
\frac{du_{\ell,2i}}{du_{e,3i}}
\left(\frac{du_e}{d\tau_e}\right)_{3i}
-1
\\[6pt]
&-\frac12\Bigg[
\left(\frac{d\tau_e}{du_e}\right)_1
\frac{du_{e,1}}{du_{\ell,2r}}
\left(\frac{du_\ell}{ds_\ell}\right)_{2r}
\left(\frac{ds_\ell}{d\tau_\ell}\right)_{2r}
\\
&\qquad\times
\left(\frac{d\tau_\ell}{ds_\ell}\right)_{2c}
\left(\frac{ds_\ell}{du_\ell}\right)_{2c}
\frac{du_{\ell,2c}}{du_{e,3c}}
\left(\frac{du_e}{d\tau_e}\right)_{3c}
-1
\Bigg].
\end{aligned}
\end{equation}

Our primary objective is to investigate the theoretical suppression of signal-propagation effects by the triple-link combination, rather than the detailed implementation of transponder hardware and link design. We therefore consider an idealized link configuration in which the uplink reception and both downlink emissions occur simultaneously at $t_2$, the two downlink signals are received simultaneously at $t_3$, and both downlinks experience identical signal-propagation effects, with any differences arising from their propagation paths, carrier characteristics, or other link conditions neglected. Accordingly,
\begin{subequations}
\label{eq:eme_assumptions}
\begin{equation}
\label{eq:eme_assumption_t2} 
t_{2,r}=t_{2,c}=t_{2,i}=t_2,
\end{equation}
\begin{equation}
\label{eq:eme_assumption_t3}
t_{3,c}=t_{3,i}=t_3,
\end{equation}
\begin{equation}
\label{eq:eme_assumption_propagation}
\frac{du_{\ell,2i}}{du_{e,3i}}=\frac{du_{\ell,2c}}{du_{e,3c}}.
\end{equation}
\end{subequations}

Under these assumptions, the triple-link observable simplifies to
\begin{equation}
\label{eq:eme_simplified}
\begin{aligned}
y_{\mathrm{tri}}
={}&
\left(\frac{d\tau_\ell}{ds_\ell}\right)_2
\left(\frac{ds_\ell}{du_\ell}\right)_2
\frac{du_{\ell,2}}{du_{e,3}}
\left(\frac{du_e}{d\tau_e}\right)_3
-1
\\[6pt]
&-\frac12\Bigg[
\left(\frac{d\tau_e}{du_e}\right)_1
\frac{du_{e,1}}{du_{\ell,2}}
\frac{du_{\ell,2}}{du_{e,3}}
\left(\frac{du_e}{d\tau_e}\right)_3
-1
\Bigg].
\end{aligned}
\end{equation}

Using Eq.~\eqref{eq:frequency_shift_simply}, the triple-link fractional-frequency observable can be expressed as
\begin{equation}
\label{eq:eme_final}
\begin{aligned}
y_{\mathrm{tri}}={}&\mathcal{F}_{1,\mathrm{tri}}+\mathcal{F}_2^{2\to3}+\mathcal{F}_1^{2\to3}\mathcal{F}_2^{2\to3} \\
&-\frac12\left(1+\mathcal{F}_1^{1\to2}+\mathcal{F}_1^{2\to3}\right)\Delta_e+O(c^{-4}).
\end{aligned}
\end{equation}

Here, $\Delta_e$ represents the change in the E-clock rate between the initial uplink emission at $t_1$ and the final downlink reception at $t_3$, defined as
\begin{equation}
\label{eq:delta_e}
\begin{aligned}
\Delta_e &=\frac{(d\tau_e/du_e)_1}{(d\tau_e/du_e)_3}-1 \\
&=\frac{W_e(t_3)-W_e(t_1)}{c^2}+O(c^{-4}).
\end{aligned}
\end{equation}

The Earth--Moon round-trip light travel time is approximately 2.5~s. If the E-clock rate variation over this interval is negligible at the required accuracy, the term proportional to $\Delta_e$ may be omitted.

The clock-rate term $\mathcal{F}_2^{2\to3}$ is associated with the Moon-to-Earth downlink and is given by \[\mathcal{F}_2^{2\to3}=\frac{W_e(t_3)-W_\ell(t_2)+F(t_2)}{c^2}.\]

The residual signal-propagation effect in the triple-link combination is
\begin{equation}
\label{eq:F1_triple}
\mathcal{F}_{1,\mathrm{tri}}=\frac12\left(\mathcal{F}_1^{2\to3}-\mathcal{F}_1^{1\to2}-\mathcal{F}_1^{1\to2}\mathcal{F}_1^{2\to3}\right),
\end{equation}
where $\mathcal{F}_1^{1\to2}$ and $\mathcal{F}_1^{2\to3}$ represent the signal-propagation effects for the uplink (Earth to Moon in this configuration) and the downlink (Moon to Earth):
\[
\mathcal{F}_1^{1\to2}=\frac{du_{e,1}}{du_{\ell,2}}-1,
\qquad
\mathcal{F}_1^{2\to3}=\frac{du_{\ell,2}}{du_{e,3}}-1.
\]

For a Moon-to-Earth leg, $\mathcal{F}_1$ can be calculated directly using Eqs.~\eqref{eqD1}--\eqref{eqD3}. For an Earth-to-Moon leg, the same expressions are used, except that the second-order Doppler term in Eq.~\eqref{eqD1} becomes
\[+\frac{(\boldsymbol{n}_{e\ell}\cdot\boldsymbol{v}_{e\ell})(\boldsymbol{n}_{e\ell}\cdot\boldsymbol{v}_e)}{c^2},\]
while the Shapiro-delay and tropospheric terms in Eqs.~\eqref{eqD2}--\eqref{eqD3} retain the same form.

For the Moon--Earth--Moon configuration, the sequence is reversed. The L-clock transmits a signal at $t_1$, which is received by the E-clock at $t_{2,r}$. The received signal is coherently retransmitted toward the Moon at $t_{2,c}$, while an independent signal generated by the E-clock is emitted at $t_{2,i}$. The coherent and independent downlink signals are received by the L-clock at $t_{3,c}$ and $t_{3,i}$, respectively. The corresponding frequencies are denoted by $f_1$, $f_{2,r}$, $f_{2,c}$, $f_{2,i}$, $f_{3,c}$, and $f_{3,i}$.

Following Eq.~\eqref{eq:frequency_shift}, the fractional-frequency shifts for the uplink (Moon to Earth), coherent downlink (Earth to Moon), and independent downlink (Earth to Moon) are  
\begin{subequations}
\label{eq:mem_three_links}
\begin{equation}
\label{eq:mem_uplink}
\begin{aligned}
\frac{f_{2,r}}{f_1}-1={}&\left(\frac{d\tau_\ell}{ds_\ell}\right)_1\left(\frac{ds_\ell}{du_\ell}\right)_1\\
&\times \frac{du_{\ell,1}}{du_{e,2r}} \left(\frac{du_e}{d\tau_e}\right)_{2r}-1,
\end{aligned}
\end{equation}
\begin{equation}
\label{eq:mem_coherent_downlink}
\begin{aligned}
\frac{f_{3,c}}{f_{2,c}}-1={}&\left(\frac{d\tau_e}{du_e}\right)_{2c}\frac{du_{e,2c}}{du_{\ell,3c}}\\
&\times\left(\frac{du_\ell}{ds_\ell}\right)_{3c}\left(\frac{ds_\ell}{d\tau_\ell}\right)_{3c}-1,
\end{aligned}
\end{equation}
\begin{equation}
\label{eq:mem_independent_downlink}
\begin{aligned}
\frac{f_{3,i}}{f_{2,i}}-1={}&\left(\frac{d\tau_e}{du_e}\right)_{2i}\frac{du_{e,2i}}{du_{\ell,3i}}\\
&\times \left(\frac{du_\ell}{ds_\ell}\right)_{3i}\left(\frac{ds_\ell}{d\tau_\ell}\right)_{3i}-1.
\end{aligned}
\end{equation}
\end{subequations}

Applying the same triple-link combination in Eq.~\eqref{eq:triple_observable} and the idealized timing and propagation assumptions in Eqs.~\eqref{eq:eme_assumption_t2}--\eqref{eq:eme_assumption_propagation}, and expanding the resulting expression to the same order as in Eq.~\eqref{eq:frequency_shift_simply}, we obtain
\begin{equation}
\label{eq:mem_final}
\begin{aligned}
y_{\mathrm{tri}}={}&\mathcal{F}_{1,\mathrm{tri}}-\mathcal{F}_2^{2\to3}-\mathcal{F}_1^{2\to3}\mathcal{F}_2^{2\to3}\\
&-\frac12\left(1+\mathcal{F}_1^{1\to2}+\mathcal{F}_1^{2\to3}\right)\Delta_\ell+O(c^{-4}).
\end{aligned}
\end{equation}

Here, $\Delta_\ell$ represents the change in the L-clock rate between the initial uplink emission at $t_1$ and the final downlink reception at $t_3$, defined as
\begin{equation}
\label{eq:delta_l}
\begin{aligned}
\Delta_\ell&=\frac{(d\tau_\ell/ds_\ell)_1(ds_\ell/du_\ell)_1}{(d\tau_\ell/ds_\ell)_3(ds_\ell/du_\ell)_3}-1\\
&=\frac{W_\ell(t_3)-W_\ell(t_1)+F(t_1)-F(t_3)}{c^2}+O(c^{-4}).
\end{aligned}
\end{equation}

This term accounts for variations in both the lunar gravity potential and the TCG--TCL coordinate-time transformation rate. As in the Earth--Moon--Earth configuration, the term proportional to $\Delta_\ell$ may be neglected if its contribution over the approximately 2.5~s round-trip light travel time is below the required accuracy.

The clock-rate term $\mathcal{F}_2^{2\to3}$ is now associated with the Earth-to-Moon downlink and is given by \[\mathcal{F}_2^{2\to3}=\frac{W_e(t_2)-W_\ell(t_3)+F(t_3)}{c^2}.\]

The signal-propagation terms $\mathcal{F}_1^{1\to2}$ and $\mathcal{F}_1^{2\to3}$ correspond to the Moon-to-Earth leg and the Earth-to-Moon leg:
\[
\mathcal{F}_1^{1\to2}=\frac{du_{\ell,1}}{du_{e,2}}-1,
\qquad
\mathcal{F}_1^{2\to3}=\frac{du_{e,2}}{du_{\ell,3}}-1.
\]
The residual signal-propagation term $\mathcal{F}_{1,\mathrm{tri}}$ retains the same form as in the Earth--Moon--Earth configuration, as given in Eq.~\eqref{eq:F1_triple}.

Compared with a single link, the triple-link combination can significantly suppresses the signal-propagation effect. As can be seen from Eqs.~\eqref{eqD1} and \eqref{eq:F1_triple}, if the line-of-sight projection of the relative velocity is identical for the uplink and downlink legs, the first-order Doppler terms cancel completely. However, during the light-travel time, both the line of sight and the endpoint velocities may change; therefore, although the first-order Doppler effect is largely cancelled, a residual remains.

For all clock-site combinations with elevations above $3^\circ$ on both physical legs, the total residual signal-propagation effect in the triple-link combination, $\mathcal{F}_{1,\mathrm{tri}}$, ranges from $-1.34\times10^{-10}$ to $4.10\times10^{-11}$ for the Earth--Moon--Earth configuration and from $-3.05\times10^{-12}$ to $5.12\times10^{-12}$ for the Moon--Earth--Moon configuration. Compared with the $10^{-6}$ level for a single link, this represents a reduction of approximately four to six orders of magnitude. Tab.~\ref{tab:triple_residual} lists the corresponding ranges for the individual E-clock sites and different elevation intervals. Unlike the single-link results, the triple-link residual does not show a clear decrease in magnitude with increasing elevation.

$\mathcal{F}_{1,\mathrm{tri}}$ is dominated by the incompletely cancelled first-order Doppler term. For elevations above $3^\circ$, the first-order Doppler residual ranges from $-1.34\times10^{-10}$ to $4.16\times10^{-11}$ for the Earth--Moon--Earth configuration and from $-3.03\times10^{-12}$ to $5.13\times10^{-12}$ for the Moon--Earth--Moon configuration. In the Earth--Moon--Earth case, the first-order Doppler residual mainly results from the change in the Earth-site velocity during the two-way propagation, primarily due to Earth rotation. In the Moon--Earth--Moon case, it mainly results from the change in the lunar-site velocity, primarily due to lunar orbital acceleration. 

These different motions (Earth rotation and lunar orbital motion) lead to different changes in the clock-site velocities during the two-way signal propagation and largely account for the different magnitudes of the residual signal-propagation term $\mathcal F_{1,\mathrm{tri}}$ in the two configurations. Based on Eqs.~\eqref{eqD1} and \eqref{eq:F1_triple}, the incompletely cancelled first-order Doppler contribution to $\mathcal F_{1,\mathrm{tri}}$ can be written as
\[
\mathcal F_{1\mathrm{dp},\mathrm{tri}}^{(1)}
=-\frac{1}{2c}
\left(
\bm n_{32}\cdot\bm v_{32}
-\bm n_{21}\cdot\bm v_{21}
\right).
\]
Here, $\bm n_{21}$ and $\bm n_{32}$ are the unit vectors along the uplink ($1\to2$) and downlink ($2\to3$), and the corresponding relative velocities are $\bm v_{21}=\bm v_2-\bm v_1$ and $\bm v_{32}=\bm v_3-\bm v_2$.

For a simplified illustration, assuming $\bm n_{32}=-\bm n_{21}$ and an approximately constant acceleration during the round-trip light travel time between the E-clock and the L-clock, we obtain
\begin{equation*}
\begin{aligned}
\mathcal F_{1\mathrm{dp},\mathrm{tri}}^{(1)}
&\simeq-\frac{1}{2c}\bm n_{32}\cdot(\bm v_3-\bm v_1)\\
&\simeq-\frac{\Delta t}{2c}\bm n_{32}\cdot\bm a,
\end{aligned}
\end{equation*}
where $\Delta t=t_3-t_1$, and $\bm a$ denotes the acceleration in the GCRS of the E-clock for the Earth--Moon--Earth configuration and the L-clock for the Moon--Earth--Moon configuration.

For the six E-clock sites considered here, the acceleration due to Earth rotation ranges from approximately $0.022$ to $0.032\ \mathrm{m\,s^{-2}}$, about one order of magnitude larger than the typical lunar orbital acceleration of approximately $0.0027\ \mathrm{m\,s^{-2}}$ for the L-clock sites. This difference in acceleration magnitudes largely explains the larger magnitude of $\mathcal F_{1,\mathrm{tri}}$ in the Earth--Moon--Earth configuration. The precise magnitude difference between the two configurations also depends on changes in the line-of-sight geometry between the uplink and downlink, as well as on second-order Doppler and other propagation effects.

For elevations above $3^\circ$, the second-order Doppler contribution ranges from $-2.39\times10^{-12}$ to $1.82\times10^{-14}$ and from $-3.44\times10^{-13}$ to $3.48\times10^{-13}$ for the two configurations, respectively; the corresponding tropospheric contributions range from $-1.31\times10^{-13}$ to $-2.93\times10^{-17}$ and from $1.05\times10^{-18}$ to $5.81\times10^{-15}$. The Shapiro-delay contribution is much smaller, remaining within $[-1.67\times10^{-19},\,1.26\times10^{-20}]$ and $[-1.05\times10^{-21},\,7.16\times10^{-21}]$, and can therefore be neglected at the $10^{-18}$ level.

\begin{table*}
\caption{\label{tab:triple_residual}
Ranges of the total signal-propagation residual after the
triple-link cancellation for the six E-clock sites over more than
50 years. For each E-clock site, the ranges are taken over all five
L-clock sites. }
\centering
\small
\renewcommand{\arraystretch}{1.15}
\setlength{\tabcolsep}{5pt}
\begin{tabular}{l|cc|cc}
\hline\hline
& \multicolumn{2}{c|}{Earth--Moon--Earth ($\times10^{-10}$)}
& \multicolumn{2}{c}{Moon--Earth--Moon ($\times10^{-12}$)} \\
E-clock
& $3^\circ<\epsilon<30^\circ$
& $\epsilon\ge30^\circ$
& $3^\circ<\epsilon<30^\circ$
& $\epsilon\ge30^\circ$ \\
\hline
LURE
& $[-0.95538,\ 0.14098]$
& $[-1.33628,\ -0.48039]$
& $[-2.08823,\ 4.36547]$
& $[0.66065,\ 5.12466]$ \\
MLRS2
& $[-1.05954,\ 0.24457]$
& $[-1.22768,\ -0.37387]$
& $[-2.43812,\ 4.73992]$
& $[0.28845,\ 4.78504]$ \\
APOLLO
& $[-1.06012,\ 0.26247]$
& $[-1.19981,\ -0.35555]$
& $[-2.55323,\ 4.69151]$
& $[0.17284,\ 4.69313]$ \\
MLRO
& $[-1.01746,\ 0.33908]$
& $[-1.08164,\ -0.28086]$
& $[-2.81047,\ 4.30541]$
& $[-0.01760,\ 4.28674]$ \\
OCA
& $[-0.98569,\ 0.36594]$
& $[-1.02938,\ -0.25340]$
& $[-2.92094,\ 4.13618]$
& $[0.00302,\ 4.08309]$ \\
WLRS
& $[-0.91323,\ 0.41026]$
& $[-0.93150,\ -0.21017]$
& $[-3.05196,\ 3.82561]$
& $[-0.15315,\ 3.76211]$ \\
\hline\hline
\end{tabular}
\end{table*}
%%%%%%%%%%%%%Simulated frequency differences%%%%%%%%%%%%%%%%%%%
\section{Simulated frequency differences}
\begin{figure*}
\captionsetup{font=small}
\includegraphics[width=\linewidth]{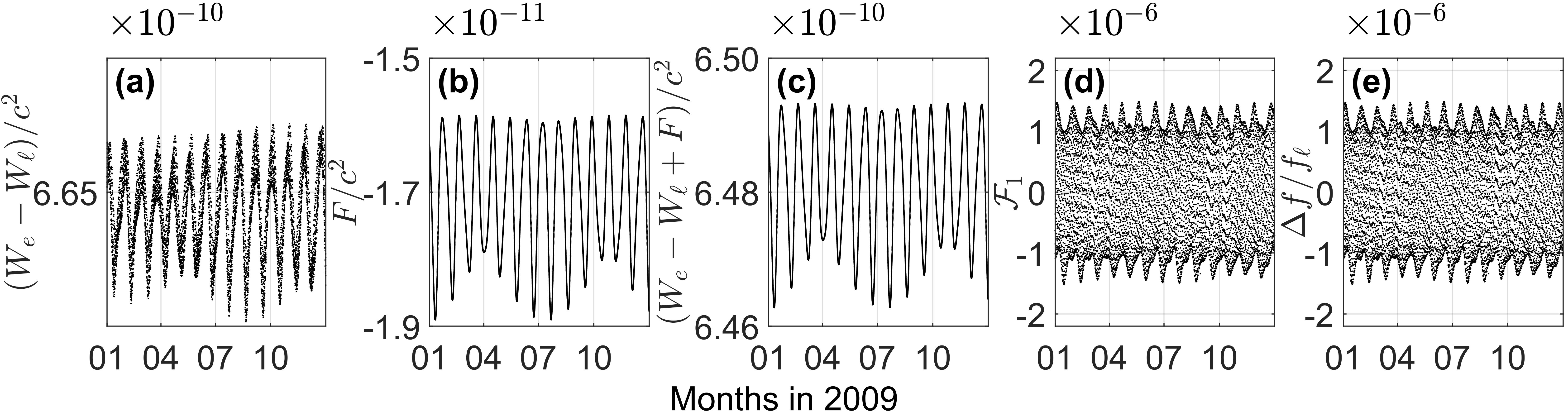}
\caption{\label{Fig3}Results in 2009 for the E-clock at APOLLO and L-clock at A15 as an example: (a) $(W_e-W_\ell)/c^2$, (b) $F/c^2$, (c) $(W_e-W_\ell+F)/c^2$, (d) $\mathcal{F}_1$ and (e) $\Delta f/f_\ell$.} % \textcolor{red}{The corresponding ranges are (a) $[6.65182,\,6.65183]\times10^{-10}$, (b) $[-1.89052,\,-1.58552]\times10^{-11}$, (c) $[-2.92459,\,2.97707]\times10^{-6}$, (d) $[6.46277,\,6.49328]\times10^{-10}$, and (e) $[-2.92394,\,2.97771]\times10^{-6}$, respectively.}
\end{figure*}

The effects on $\Delta f/f_\ell$ from the gravity potentials of E-clocks $W_e/c^2$ and L-clocks $W_\ell/c^2$ are on the order of $10^{-10}$ and $10^{-11}$, respectively. Their differences $(W_e-W_\ell)/c^2$ are on the $10^{-10}$ level, as illustrated by the example in Fig.~\ref{Fig3}(a). Results for $F/c^2$ from Eqs.~\eqref{eq8} and \eqref{eq9} are highly consistent, both indicating values at the $10^{-11}$ level (see Fig.~\ref{Fig3}(b)), with the detailed ranges already shown in Tab.~\ref{tab:F_c2}. Combining these terms, the frequency comparison excluding the signal-propagation effect, $\mathcal{F}_2=(W_e-W_\ell+F)/c^2$, remains at the $10^{-10}$ level, as illustrated in Fig.~\ref{Fig3}(c). The mean values and ranges of $(W_e-W_\ell)/c^2$ and $\mathcal{F}_2$ for all simulated Earth--Moon clock pairs are summarized in Tab.~\ref{tab:clock_rate_pairs}, where $\mathcal{F}_2$ is also expressed as the equivalent accumulated time difference in $\mu\mathrm{s/day}$. For a single link, the signal-propagation term $\mathcal{F}_1=\mathcal{F}_{1dp}+\mathcal{F}_{1sha}+\mathcal{F}_{1trop}$, illustrated in Fig.~\ref{Fig3}(d), is dominated by the first-order Doppler contribution at the $10^{-6}$ level. The corresponding station- and elevation-dependent ranges are summarized in Tab.~\ref{tab:F1_components}. Including the signal-propagation effect, the fractional frequency difference $\Delta f/f_\ell$ for the single link reaches the $10^{-6}$ level, as illustrated in Fig.~\ref{Fig3}(e), with the ranges for all simulated Earth--Moon clock pairs summarized in Tab.~\ref{tab:dff_range}.

\begin{table*}
\caption{\label{tab:clock_rate_pairs}
Mean values and ranges of $(W_e-W_\ell)/c^2$ and
$\mathcal{F}_2=(W_e-W_\ell+F)/c^2$ for all simulated Earth--Moon clock
pairs over the common period from April 1, 2002 to June 30, 2017,
determined by the availability of the daily ITSG-Grace2018 model
used for the non-tidal component of $W_e$.
$\mathcal{F}_2$ is given both as a contribution to the fractional frequency
difference and as the equivalent accumulated time difference per
day in $\mu\mathrm{s/day}$.}
\small
\renewcommand{\arraystretch}{1.15}
\setlength{\tabcolsep}{3pt}
\begin{ruledtabular}
\begin{tabular}{l|cc|cc|cc}
\multicolumn{1}{l|}{}
&
\multicolumn{2}{c|}{$(W_e-W_\ell)/c^2$ ($\times10^{-10}$)}
&
\multicolumn{2}{c|}{$\mathcal{F}_2=(W_e-W_\ell+F)/c^2$ ($\times10^{-10}$)}
&
\multicolumn{2}{c}{$\mathcal{F}_2$ ($\mu\mathrm{s/day}$)}
\\
\multicolumn{1}{l|}{E--L pair}
& Mean & Range
& Mean & Range
& Mean & Range
\\
\hline
LURE--A11
& 6.6516017 & [6.6516008, 6.6516024]
& 6.4806660 & [6.4618421, 6.4931599]
& 55.9929540 & [55.8303150, 56.1009020]
\\
LURE--A14
& 6.6517635 & [6.6517626, 6.6517643]
& 6.4808278 & [6.4620063, 6.4933259]
& 55.9943520 & [55.8317340, 56.1023360]
\\
LURE--A15
& 6.6516086 & [6.6516075, 6.6516094]
& 6.4806729 & [6.4618201, 6.4931834]
& 55.9930130 & [55.8301260, 56.1011050]
\\
LURE--L17
& 6.6515330 & [6.6515321, 6.6515339]
& 6.4805973 & [6.4617412, 6.4931117]
& 55.9923600 & [55.8294440, 56.1004850]
\\
LURE--L21
& 6.6514695 & [6.6514685, 6.6514703]
& 6.4805337 & [6.4616869, 6.4930421]
& 55.9918110 & [55.8289750, 56.0998840]
\\
\hline
MLRS2--A11
& 6.6527136 & [6.6527127, 6.6527143]
& 6.4817779 & [6.4629535, 6.4942721]
& 56.0025610 & [55.8399180, 56.1105110]
\\
MLRS2--A14
& 6.6528754 & [6.6528745, 6.6528762]
& 6.4819397 & [6.4631177, 6.4944380]
& 56.0039590 & [55.8413370, 56.1119450]
\\
MLRS2--A15
& 6.6527205 & [6.6527195, 6.6527214]
& 6.4817848 & [6.4629315, 6.4942954]
& 56.0026200 & [55.8397280, 56.1107120]
\\
MLRS2--L17
& 6.6526449 & [6.6526440, 6.6526458]
& 6.4817092 & [6.4628526, 6.4942237]
& 56.0019670 & [55.8390470, 56.1100930]
\\
MLRS2--L21
& 6.6525814 & [6.6525804, 6.6525823]
& 6.4816456 & [6.4627983, 6.4941541]
& 56.0014180 & [55.8385780, 56.1094920]
\\
\hline
APOLLO--A11
& 6.6518640 & [6.6518631, 6.6518647]
& 6.4809282 & [6.4621038, 6.4934224]
& 55.9952200 & [55.8325770, 56.1031690]
\\
APOLLO--A14
& 6.6520258 & [6.6520249, 6.6520265]
& 6.4810901 & [6.4622680, 6.4935884]
& 55.9966180 & [55.8339960, 56.1046040]
\\
APOLLO--A15
& 6.6518709 & [6.6518698, 6.6518717]
& 6.4809351 & [6.4620819, 6.4934458]
& 55.9952790 & [55.8323880, 56.1033710]
\\
APOLLO--L17
& 6.6517953 & [6.6517944, 6.6517961]
& 6.4808595 & [6.4620030, 6.4933740]
& 55.9946260 & [55.8317060, 56.1027520]
\\
APOLLO--L21
& 6.6517317 & [6.6517308, 6.6517326]
& 6.4807960 & [6.4619487, 6.4933045]
& 55.9940770 & [55.8312370, 56.1021510]
\\
\hline
MLRO--A11
& 6.6543867 & [6.6543859, 6.6543875]
& 6.4834510 & [6.4646267, 6.4959451]
& 56.0170170 & [55.8543750, 56.1249660]
\\
MLRO--A14
& 6.6545485 & [6.6545477, 6.6545493]
& 6.4836128 & [6.4647909, 6.4961111]
& 56.0184150 & [55.8557940, 56.1264000]
\\
MLRO--A15
& 6.6543936 & [6.6543926, 6.6543945]
& 6.4834579 & [6.4646048, 6.4959684]
& 56.0170760 & [55.8541850, 56.1251670]
\\
MLRO--L17
& 6.6543180 & [6.6543172, 6.6543189]
& 6.4833823 & [6.4645259, 6.4958966]
& 56.0164230 & [55.8535040, 56.1245470]
\\
MLRO--L21
& 6.6542545 & [6.6542536, 6.6542554]
& 6.4833188 & [6.4644716, 6.4958271]
& 56.0158740 & [55.8530350, 56.1239460]
\\
\hline
OCA--A11
& 6.6535351 & [6.6535342, 6.6535358]
& 6.4825993 & [6.4637751, 6.4950935]
& 56.0096580 & [55.8470160, 56.1176070]
\\
OCA--A14
& 6.6536969 & [6.6536960, 6.6536976]
& 6.4827612 & [6.4639393, 6.4952594]
& 56.0110560 & [55.8484350, 56.1190420]
\\
OCA--A15
& 6.6535420 & [6.6535410, 6.6535428]
& 6.4826062 & [6.4637531, 6.4951167]
& 56.0097180 & [55.8468270, 56.1178080]
\\
OCA--L17
& 6.6534664 & [6.6534655, 6.6534673]
& 6.4825306 & [6.4636742, 6.4950449]
& 56.0090640 & [55.8461450, 56.1171880]
\\
OCA--L21
& 6.6534028 & [6.6534019, 6.6534037]
& 6.4824671 & [6.4636199, 6.4949754]
& 56.0085160 & [55.8456760, 56.1165870]
\\
\hline
WLRS--A11
& 6.6542476 & [6.6542469, 6.6542484]
& 6.4833119 & [6.4644876, 6.4958060]
& 56.0158150 & [55.8531730, 56.1237640]
\\
WLRS--A14
& 6.6544094 & [6.6544086, 6.6544102]
& 6.4834737 & [6.4646518, 6.4959720]
& 56.0172130 & [55.8545920, 56.1251980]
\\
WLRS--A15
& 6.6542545 & [6.6542536, 6.6542554]
& 6.4833188 & [6.4644657, 6.4958293]
& 56.0158740 & [55.8529830, 56.1239650]
\\
WLRS--L17
& 6.6541789 & [6.6541781, 6.6541798]
& 6.4832432 & [6.4643868, 6.4957575]
& 56.0152210 & [55.8523020, 56.1233450]
\\
WLRS--L21
& 6.6541154 & [6.6541145, 6.6541162]
& 6.4831797 & [6.4643325, 6.4956880]
& 56.0146720 & [55.8518330, 56.1227440]
\\
\end{tabular}
\end{ruledtabular}
\end{table*}

\begin{table}
\caption{\label{tab:dff_range}
Ranges of the simulated fractional frequency difference
$\Delta f/f_{\ell}$ for all assumed Earth--Moon clock pairs
over the common period from April 1, 2002 to June 30, 2017.
Only epochs with elevation angles $\epsilon>3^\circ$ are included.}
\begin{ruledtabular}
\begin{tabular}{lc}
E--L pair & $\Delta f/f_{\ell}$ ($\times10^{-6}$) \\
\hline
LURE--A11   & $[-1.64684,\ 1.64736]$ \\
LURE--A14   & $[-1.64652,\ 1.64780]$ \\
LURE--A15   & $[-1.64613,\ 1.64685]$ \\
LURE--L17   & $[-1.64560,\ 1.64684]$ \\
LURE--L21   & $[-1.64637,\ 1.64669]$ \\

MLRS2--A11  & $[-1.53669,\ 1.53612]$ \\
MLRS2--A14  & $[-1.53745,\ 1.53661]$ \\
MLRS2--A15  & $[-1.53609,\ 1.53561]$ \\
MLRS2--L17  & $[-1.53608,\ 1.53563]$ \\
MLRS2--L21  & $[-1.53572,\ 1.53542]$ \\

APOLLO--A11 & $[-1.50712,\ 1.51205]$ \\
APOLLO--A14 & $[-1.50776,\ 1.51242]$ \\
APOLLO--A15 & $[-1.50643,\ 1.51152]$ \\
APOLLO--L17 & $[-1.50633,\ 1.51146]$ \\
APOLLO--L21 & $[-1.50613,\ 1.51139]$ \\

MLRO--A11   & $[-1.38883,\ 1.39661]$ \\
MLRO--A14   & $[-1.38881,\ 1.39631]$ \\
MLRO--A15   & $[-1.38804,\ 1.39577]$ \\
MLRO--L17   & $[-1.38760,\ 1.39521]$ \\
MLRO--L21   & $[-1.38812,\ 1.39601]$ \\

OCA--A11    & $[-1.32691,\ 1.33981]$ \\
OCA--A14    & $[-1.32757,\ 1.33977]$ \\
OCA--A15    & $[-1.32611,\ 1.33903]$ \\
OCA--L17    & $[-1.32598,\ 1.33864]$ \\
OCA--L21    & $[-1.32602,\ 1.33914]$ \\

WLRS--A11   & $[-1.23533,\ 1.23933]$ \\
WLRS--A14   & $[-1.23552,\ 1.23926]$ \\
WLRS--A15   & $[-1.23456,\ 1.23857]$ \\
WLRS--L17   & $[-1.23423,\ 1.23817]$ \\
WLRS--L21   & $[-1.23452,\ 1.23869]$ \\
\end{tabular}
\end{ruledtabular}
\end{table}

Based on the models used and clock sites assumed in this work, the uncertainties in $\mathcal{F}_2$ arising from different sources are summarized in Tab.~\ref{tab:F2_uncertainty}.

For the Earth static gravitational field, the uncertainties of the EIGEN-6C4 coefficients propagate into $\mathcal{F}_2$ through $V_{E\text{st},e}/c^2$. As discussed in Sec.~\ref{sec:grav-pot-E} and Appendix~\ref{appB}, EIGEN-6C4 is used through degree and order 2190. Among the assumed E-clock sites, the $\mathcal{F}_2$ uncertainty is largest at WLRS, reaching $6.8\times10^{-18}$ (see also Tab.~\ref{Tab4}).

The uncertainties of the E-clock site coordinates propagate into $\mathcal{F}_2$ through $W_e/c^2$, mainly through the static gravitational potential $V_{E\text{st},e}/c^2$. The resulting $\mathcal{F}_2$ uncertainty reaches approximately $2.1\times10^{-18}$ at WLRS and $1.4\times10^{-18}$ at MLRO, while remaining below $1\times10^{-18}$ at the other four E-clock sites. The larger values at WLRS and MLRO result from their comparatively larger coordinate uncertainties in the present LLR solution (see Tab.~\ref{tab:station_coordinates}), which in turn are due to the smaller number of accumulated LLR observations at these stations.

For the lunar static gravitational field, the uncertainties of the GL0900D coefficients propagate into $\mathcal{F}_2$ through $V_{L\text{st},\ell}/c^2$. As discussed in Sec.~\ref{sec:grav-pot-L} and Appendix~\ref{appB}, the coefficients are retained through degree 500. Among the assumed L-clock sites, the $\mathcal{F}_2$ uncertainty from the retained coefficients is largest at L21, reaching $2.1\times10^{-19}$ (see also Tab.~\ref{Tab6}).

The uncertainties of the L-clock site coordinates propagate into $\mathcal{F}_2$ through both $W_\ell/c^2$ and $F/c^2$, predominantly through the monopole term $GM_L/r$ of the lunar static gravitational potential. Among the assumed L-clock sites, the $\mathcal{F}_2$ uncertainty is largest at A14, reaching approximately $4.4\times10^{-19}$.

Celestial-body position uncertainties affect $\mathcal{F}_2$ mainly through $F/c^2$. As evaluated in Appendix~\ref{appC}, the resulting uncertainties are no larger than $2.6\times10^{-19}$ over more than 50 years, with the dominant contribution arising from the uncertainty in the Sun’s position.

The uncertainties of the lunar orientation parameters (the three lunar Euler angles \cite{Folkner2014}) and angular velocity propagate into $\mathcal{F}_2$ through $F/c^2$, the lunar tidal potential, and the lunar centrifugal potential. The combined propagated uncertainty in $\mathcal{F}_2$ from the lunar orientation parameters and angular velocity does not exceed $2.3\times10^{-22}$ for the assumed L-clock sites over 1970–2023.

The effect of length of day (LOD) is also considered in the Earth centrifugal potential. Based on the uncertainties of the IERS EOP C04 LOD series, the resulting uncertainty in $\mathcal{F}_2$ does not exceed $3.5\times10^{-20}$ for the assumed E-clock sites over the series period 1962-2025.

For the single-link configuration, the signal-propagation term $\mathcal{F}_1$ provides the dominant contribution to the fractional frequency difference between the Earth and Moon clocks, reaching the $10^{-6}$ level and being dominated by the first-order Doppler effect. Because this term is much larger than the other contributions and is highly sensitive to the relative geometry and motion of the link endpoints, even small uncertainties in the parameters entering $\mathcal{F}_1$ can produce comparatively large propagated uncertainties. The propagated uncertainties in $\mathcal{F}_1$ from the considered uncertainty sources are summarized in Tab.~\ref{tab:F1_uncertainty}.

The uncertainties of the assumed E-clock and L-clock site coordinates propagate into $\mathcal{F}_1$ mainly through the Doppler term. Among the assumed sites, the maximum propagated uncertainties in $\mathcal{F}_1$ for the single-link configuration are approximately $2.6\times10^{-15}$ for the E-clock coordinates, obtained at WLRS, and $8.0\times10^{-17}$ for the L-clock coordinates, obtained at A14.

The uncertainties of the Earth orientation parameters, including the terrestrial pole coordinates $x_p$ and $y_p$, the Earth rotation phase $UT1–UTC$, and the celestial pole offsets $dX$ and $dY$, together with the uncertainty in the Earth angular velocity derived from LOD, affect $\mathcal{F}_1$ through the GCRS position and velocity of the E-clock. Using the uncertainties from the IERS EOP C04 series, the combined propagated uncertainty from these six quantities reaches $2.0\times10^{-13}$ over 1970–2023. For the more recent period 2000–2023, the corresponding maximum decreases to $4.0\times10^{-14}$. In both periods, the uncertainty in $UT1–UTC$ is the dominant source.  

The uncertainties of the lunar orbital initial position and velocity (with respect to the Earth), lunar orientation parameters, and lunar angular velocity are specified at 00:00:00 on January 1, 2000, and their effects at other epochs are evaluated from the corresponding perturbed lunar ephemerides. Over 1970–2023, the uncertainties in the lunar orbital initial position and velocity produce an $\mathcal{F}_1$ uncertainty that is increasingly amplified as the lunar ephemeris is propagated farther from the initial epoch. Consistent with this behavior, the largest $\mathcal{F}_1$ uncertainty over the evaluated interval occurs near 1970, the part of the interval farthest from the initial epoch. The uncertainty is at the $10^{-17}$ level near the initial epoch and reaches the $10^{-12}$ level over the evaluated interval. In contrast, the $\mathcal{F}_1$ uncertainties associated with the lunar orientation parameters and angular velocity show no comparable long-term amplification, with values generally at the $10^{-17}$ level over 1970–2023.

Based on the above uncertainty analyses of $\mathcal{F}_1$ and $\mathcal{F}_2$, over the long term, the lunar orbital uncertainty becomes the dominant uncertainty source in $\Delta f/f_\ell$ for a single-link Earth–Moon clock comparison. For $\mathcal{F}_1$, the principal approach to suppressing the propagated uncertainties considered here is the use of the Earth–Moon–Earth and Moon–Earth–Moon triple-link configurations. Near the initial epoch, the lunar-orbit-induced $\mathcal{F}_1$ uncertainty is reduced from the $10^{-17}$ level for a single link to about $10^{-21}$ by both configurations. Over 1970–2023, its largest value reaches the $10^{-12}$ level for a single link, while the corresponding largest values are reduced to the $10^{-16}$ level for the Earth–Moon–Earth configuration and to a few $10^{-18}$ for the Moon–Earth–Moon configuration (see Tab.~\ref{tab:F1_uncertainty}). Continued improvements in the quality and quantity of LLR observations are also expected to further improve lunar-orbit determination and reduce the associated propagated uncertainty.

For $\mathcal{F}_2$, the dominant limitations arise instead from the gravity-field-related terms. As shown in Secs.~\ref{sec:grav-pot-E} and~\ref{sec:grav-pot-L} and Appendix~\ref{appB}, the present global Earth static gravity-field model alone is not sufficient for $10^{-18}$-level Earth–Moon clock comparisons. For the lunar static gravitational field, the uncertainty reported in Tab.~\ref{tab:F2_uncertainty} refers only to the GL0900D coefficients retained through degree 500, while the noise-dominated higher-degree field remains an issue for future high-accuracy applications. For both, additional local or site-specific gravity information may provide further improvement.

The E-clock and L-clock site-coordinate uncertainties affect both $\mathcal{F}_1$ and $\mathcal{F}_2$. For the assumed clock sites, the corresponding uncertainties propagated into $\mathcal{F}_1$ are reduced below $10^{-18}$ by the triple-link configurations; for $\mathcal{F}_2$, the uncertainty propagated from the L-clock coordinates remains below $10^{-18}$, while that from the E-clock coordinates at some sites can be further reduced as more and higher-precision LLR observations improve the station coordinates.

The remaining uncertainty sources considered for $\mathcal{F}_1$ can likewise be effectively suppressed by the triple-link configurations, whereas the remaining sources considered for $\mathcal{F}_2$ produce propagated uncertainties well below $10^{-18}$.

\begin{table*}
\centering
\captionsetup{font=small}
\caption{\label{tab:F2_uncertainty}
Largest propagated uncertainties in $\mathcal{F}_2$ from the considered uncertainty sources for the assumed clock sites and adopted models.}
\setlength{\tabcolsep}{5pt}
\renewcommand{\arraystretch}{1}
\begin{tabular}{l c l}
\hline\hline
\multicolumn{1}{c}{Uncertainty source} & Largest propagated uncertainty in $\mathcal{F}_2$ & \multicolumn{1}{c}{Remark} \\
\hline
\noalign{\vskip 3pt}
\parbox[c]{0.30\textwidth}{\centering Earth static gravity field model coefficients} & $6.8\times10^{-18}$ & \parbox[c]{0.38\textwidth}{\centering EIGEN-6C4 through degree/order 2190; largest value obtained at WLRS.}
\\ [11pt]
\parbox[c]{0.30\textwidth}{\centering E-clock site coordinates} & $2.1\times10^{-18}$ & \parbox[c]{0.38\textwidth}{\centering Largest value obtained at WLRS.}
\\ [10pt]
\parbox[c]{0.30\textwidth}{\centering Lunar static gravity field model coefficients} & $2.1\times10^{-19}$ & \parbox[c]{0.38\textwidth}{\centering GL0900D through degree 500; largest value obtained at L21. Higher degrees are excluded because of increasing noise.}
\\[18pt]
\parbox[c]{0.30\textwidth}{\centering L-clock site coordinates} & $4.4\times10^{-19}$ & \parbox[c]{0.38\textwidth}{\centering Largest value obtained at A14.}
\\[8pt]
\parbox[c]{0.30\textwidth}{\centering Celestial body positions} & $2.6\times10^{-19}$ & \parbox[c]{0.38\textwidth}{\centering Evaluated over more than 50 years; dominated by the Sun position uncertainty.}
\\[13pt]
\parbox[c]{0.30\textwidth}{\centering Earth angular velocity from LOD} & $3.5\times10^{-20}$ & \parbox[c]{0.38\textwidth}{\centering IERS EOP C04 LOD uncertainties over 1962--2025; largest value obtained at LURE.}
\\[13pt]
\parbox[c]{0.30\textwidth}{\centering Lunar orientation and angular velocity} & $2.3\times10^{-22}$ & \parbox[c]{0.38\textwidth}{\centering Evaluated over 1970--2023; largest value obtained at L17.}
\\[5pt]
\hline\hline
\end{tabular}
\end{table*}

\begin{table*}
\centering
\captionsetup{font=small}
\caption{\label{tab:F1_uncertainty}
Largest propagated uncertainties in $\mathcal{F}_1$ from the considered uncertainty sources for the assumed clock sites in the single-link and triple-link configurations, using only links with elevation angles $\epsilon>3^\circ$, evaluated at the epochs or over the intervals specified in the remarks, where applicable.}
\setlength{\tabcolsep}{5pt}
\renewcommand{\arraystretch}{1}
\begin{tabular}{l c c c l}
\hline\hline
\multicolumn{1}{c}{Uncertainty source}
& Single link
& \shortstack{Triple link:\\Earth--Moon--Earth}
& \shortstack{Triple link:\\Moon--Earth--Moon}
& \multicolumn{1}{c}{Remark}
\\
\hline
\noalign{\vskip 3pt}
\parbox[c]{0.19\textwidth}{\centering E-clock site coordinates}
& $2.6\times10^{-15}$
& $2.5\times10^{-19}$
& $1.0\times10^{-20}$
& \parbox[c]{0.33\textwidth}{\centering Largest values obtained at WLRS.}
\\[8pt]
\parbox[c]{0.19\textwidth}{\centering L-clock site coordinates}
& $8.0\times10^{-17}$
& $6.5\times10^{-21}$
& $3.2\times10^{-22}$
& \parbox[c]{0.33\textwidth}{\centering Largest values obtained at A14.}
\\[11pt]
\multirow{2}{*}{
\parbox[c]{0.19\textwidth}{\centering
Earth orientation parameters and Earth angular velocity}
}
& $2.0\times10^{-13}$
& $1.8\times10^{-17}$
& $7.2\times10^{-19}$
& \parbox[c]{0.33\textwidth}{\centering
Evaluated over 1970--2023 using uncertainties from IERS EOP C04 series.}
\\[7pt]
& $4.0\times10^{-14}$
& $2.8\times10^{-18}$
& $1.0\times10^{-19}$
& \parbox[c]{0.33\textwidth}{\centering
Evaluated over 2000--2023 using uncertainties from IERS EOP C04 series.}
\\[13pt]
\parbox[c]{0.19\textwidth}{\centering
Lunar orientation and angular velocity}
& $7.2\times10^{-17}$
& $5.7\times10^{-21}$
& $2.7\times10^{-22}$
& \parbox[c]{0.33\textwidth}{\centering
Evaluated over 1970--2023.}
\\[11pt]
\multirow{2}{*}{
\parbox[c]{0.19\textwidth}{\centering
Lunar orbital initial position and velocity}
}
& $6.2\times10^{-17}$
& $4.8\times10^{-21}$
& $1.9\times10^{-21}$
& \parbox[c]{0.33\textwidth}{\centering
Obtained near the initial epoch of 00:00:00 on January 1, 2000.}
\\[6pt]
& $1.3\times10^{-12}$
& $1.2\times10^{-16}$
& $5.6\times10^{-18}$
& \parbox[c]{0.33\textwidth}{\centering
Largest values obtained over 1970--2023.}
\\[3pt]
\hline\hline
\end{tabular}
\end{table*}

%%%%%%%%%%%%%Conclusions and outlooks%%%%%%%%%%%%%%%%%%%
\section{Conclusions and outlook}
As lunar clocks gain importance for lunar timekeeping and chronometric geodesy, particularly given navigation plans, modeling the Earth–Moon frequency link is essential. We simulated fractional frequency differences $\Delta f/f_\ell$ between E- and L-clocks, using four time transformations: the proper-to-coordinate time conversions for each clock (linked to their local gravity potentials $W_e$ and $W_\ell$), Earth--Moon coordinate time transformation and signal propagation between E- and L-clocks.

The individual contributions of $W_e/c^2$ and $W_\ell/c^2$ to $\Delta f/f_\ell$ are at the $10^{-10}$ and $10^{-11}$ levels, respectively, and their difference $(W_e-W_\ell)/c^2$ reaches the $10^{-10}$ level. Earth static gravitational potential dominates $W_e$. For the $10^{-18}$-level Earth--Moon clock frequency comparison, other $W_e$ components should also be considered: the centrifugal potential, solid Earth tide, ocean-tide loading, solid pole tide, and non-tidal signals. Lunar static gravitational potential dominates $W_\ell$, while the lunar centrifugal potential and solid tide should also be accounted for. Current global Earth and lunar gravity field models may not be sufficient by themselves to provide a fully site-specific error floor for the $10^{-18}$-level Earth--Moon clock comparison. Augmenting them with local or site-specific gravity information is therefore likely required for such high-accuracy applications. 

Unlike single-system comparisons (e.g., between E-clocks), inter-system comparisons (e.g., between E- and L-clocks) must consider the effect of the coordinate time ratio $F/c^2$. Here, it is at the $10^{-11}$ level, dominated by Earth, Moon, and Sun, with smaller effects from Jupiter, Venus, and other planets. Two methods for calculating $F/c^2$ agree at the $10^{-18}$ level, mutually validating the result. 

The Doppler, atmospheric, and Shapiro-delay terms in signal propagation are quantified. For a single Earth--Moon link, the first-order Doppler effect dominates the frequency comparison at the $10^{-6}$ level, masking the $10^{-10}$--$10^{-11}$-level gravitational-potential and coordinate-time terms. Extracting these terms while suppressing the signal-propagation effect requires a Doppler-cancelling multi-link strategy. The multi-link combination reduces the signal-propagation effect by approximately four to six orders of magnitude compared with the single-link case, so that the clock-rate term of interest, $\mathcal{F}_2=(W_e-W_\ell+F)/c^2$, is no longer overwhelmed by the signal-propagation effect.

Future Earth–Moon clock comparisons also face practical implementation issues. At the $10^{-18}$ level considered here, the corresponding gravitational-potential sensitivity is about $0.1\,\mathrm{m}^2/\mathrm{s}^2$, which roughly corresponds to a 6 cm change in the radial coordinate near the lunar surface, highlighting the importance of accurately determining the position of a lunar clock site. Decades of LLR observations have enabled high-precision determination of the positions of the long-observed Apollo and Lunokhod retroreflectors on the Moon \cite{Williams2021}, and the next-generation lunar laser retroreflector developments are aimed at further improving lunar ranging precision \cite{Currie2013,Muccino2025}. Together, these capabilities provide a promising synergy with future lunar clocks: LLR can constrain the position and temporal variations of a reflector site on the lunar surface, while a co-located clock can probe the local gravitational potential at the same point, thereby providing a common lunar fiducial with both geometric and gravitational-potential information. For lunar clock operation, autonomy and robustness are relevant, particularly when continuous on-site intervention is unavailable. Thermal control is also important for clock performance; for example, ambient thermal radiation can induce blackbody-radiation-related frequency shifts in optical clocks \cite{Hassan2025}. Beyond the clock itself, high-precision time/frequency transfer has been demonstrated over long-distance optical-fiber and free-space links \cite{Predehl2012, Shen2022}. Extending such capabilities to an Earth–Moon link is another important aspect of future clock comparisons.

Beyond these implementation aspects, the present Earth–Moon clock-comparison framework may also provide a general-relativistic baseline for future tests of fundamental physics. Possible violations of Local Position Invariance (LPI) or Local Lorentz Invariance (LLI), as well as possible variations of fundamental constants, could in principle produce anomalous signatures relative to the modeled relativistic frequency-comparison observable \cite{Will2014, Guerlin2015, Delva2018, Herrmann2018, Savalle2019, Bize2004, Sanner2019, Blatt2008, Lange2021}. However, dedicated tests of these effects would require additional specifications beyond the present work, such as specific clock species, noise models, mission design, and parameter-estimation frameworks.
%%%%%%%%%%%%%%Acknowledgments%%%%%%%%%%%%%%%%%%%
\begin{acknowledgments}
The authors acknowledge the Deutsche Forschungsgemeinschaft (DFG, German Research Foundation) under Germany’s Excellence Strategy-EXC-2123 QuantumFrontiers, Project-ID 390837967 and the SFB 1464 TerraQ, Project-ID 434617780. We also thank the anonymous referees for their constructive comments and suggestions, which helped improve the organization, clarity, and presentation of the manuscript.
\end{acknowledgments}
%%%%%%%%%%%%%Appendices%%%%%%%%%%%%%%%%%%%
\appendix
\renewcommand{\thesection}{\Alph{section}}

\makeatletter
\renewcommand{\@seccntformat}[1]{%
  \ifnum\pdfstrcmp{#1}{section}=0
    Appendix~\csname the#1\endcsname.\quad
  \else
    \csname the#1\endcsname.\quad
  \fi}
\makeatother
%%%%%%%%%%%%%Appendix A%%%%%%%%%%%%%%%%%%%
\section{Simulation flowchart}
\label{appA}
\begin{figure*}
\captionsetup{font=small}
\includegraphics[width=\linewidth]{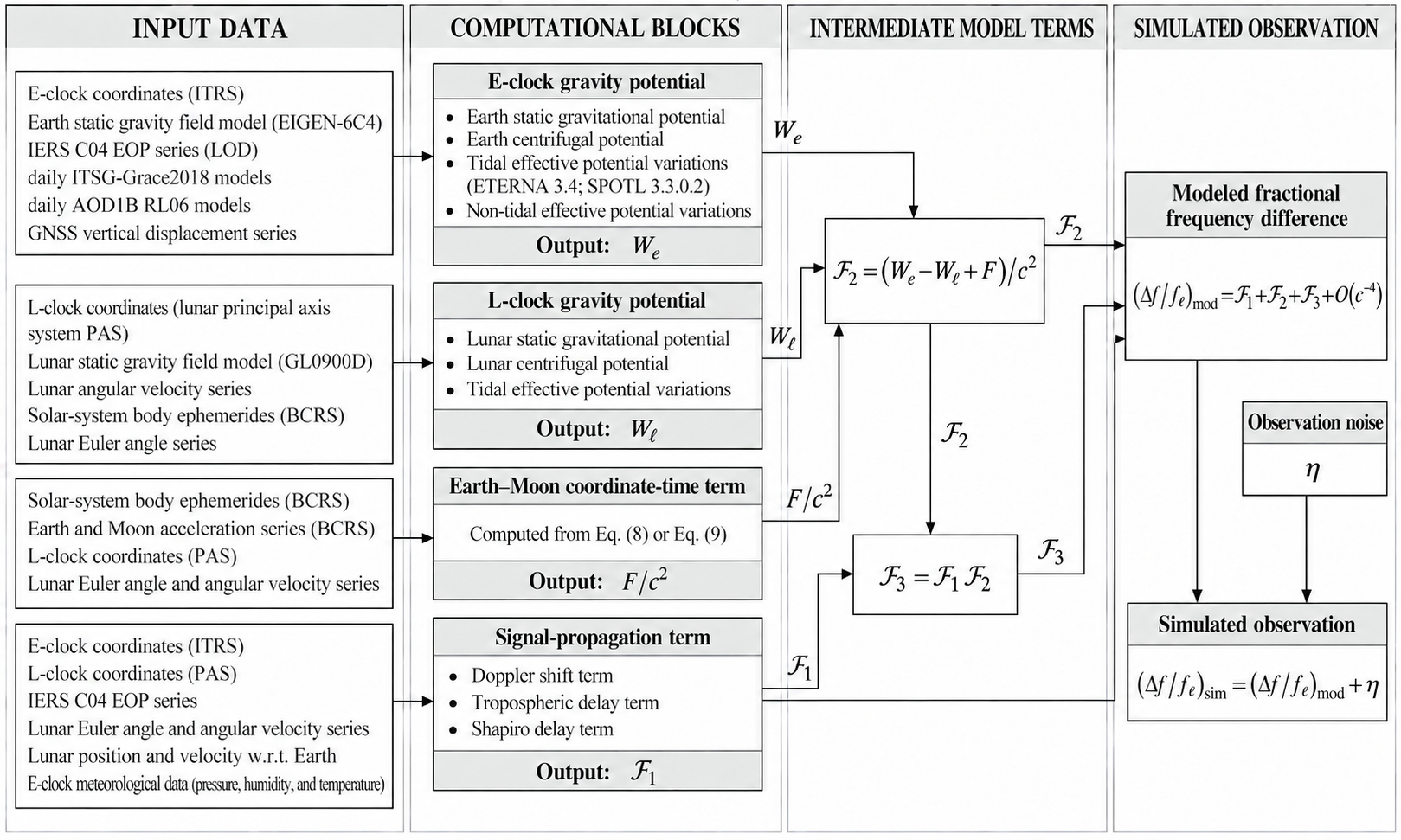}
\caption{\label{Fig4} Simulation flowchart.}
\end{figure*}

Fig.~\ref{Fig4} illustrates the simulation workflow. The E-clock and L-clock gravity potentials $W_e$ and $W_{\ell}$, the Earth--Moon coordinate-time term $F/c^2$, and the signal-propagation term $\mathcal{F}_1$ are first computed from the corresponding input data. These quantities are then combined to obtain $\mathcal{F}_2=(W_e-W_{\ell}+F)/c^2$ and $\mathcal{F}_3=\mathcal{F}_1\mathcal{F}_2$. Finally, the modeled fractional frequency difference is formed according to  Eq.~\eqref{eq:frequency_shift_simply}, and the simulated observation is obtained by adding the observation noise $\eta$.
%%%%%%%%%%%%%Appendix B%%%%%%%%%%%%%%%%%%%
\section{Earth and Moon gravity field models}
\label{appB}
\subsection{Gravity field models for the Earth}
Earth gravity field models EIGEN-6C4 (EIG)\cite{Foerste2014}, XGM2019e\_2159 (XGM) \cite{Zingerle2019} and SGG-UGM-2 (SGG) \cite{Liang2020} were compared for $V_{E\text{st},e}/c^{2}$ at six E-clocks; all provide normalized Stokes coefficients up to degree 2190 and were unified to the tide-free system. Tab.~\ref{Tab4} lists $V_{E\text{st},e}/c^{2}$ uncertainties inferred from the published coefficient uncertainties (to degree 2190) of these models. Additionally, lower-degree models--ITSG-Grace2018s (max $n=200$) \cite{MayerGuerr2018, Kvas2019} and GOCO2025s (max $n=300$) \cite{Oehlinger2025}--yielded larger uncertainties. Tab.~\ref{Tab5} gives the inter-model differences in $V_{Est,e}/c^{2}$.
\begin{table}
\captionsetup{font=small}
\caption{\label{Tab4}$V_{E\text{st},e}/c^{2}$ uncertainties at E-clocks from three Earth models, degrees up to 2190. }
\begin{ruledtabular}
\begin{tabular}{ccccccc}
Model & LURE & WLRS & OCA & MLRS2 & APOLLO & MLRO \\
& \multicolumn{6}{c}{all entries $\times 10^{-18}$} \\
\midrule
EIG & 3.8 & 6.8 & 5.6 & 4.4 & 4.4 & 5.6 \\
XGM   & 3.5 & 15.4 & 9.3 & 4.8 & 4.7 & 8.9 \\
SGG   & 2.1 & 17.5 & 8.9 & 3.2 & 3.1 & 8.2 \\
\end{tabular}
\end{ruledtabular}
\end{table}

\begin{table}
\captionsetup{font=small}
\caption{\label{Tab5}Inter-model differences in $V_{E\text{st},e}/c^{2}$ at E-clocks.}
\begin{ruledtabular}
\begin{tabular}{ccccccc}
Pair & LURE & WLRS & OCA & MLRS2 & APOLLO & MLRO \\
\multicolumn{1}{c}{} & \multicolumn{6}{c}{all entries $\times 10^{-18}$}\\
\midrule
EIG-XGM & -6.2 & -0.6 & -1.8 & -1.0 & -0.6 & 1.4 \\
EIG-SGG & -6.8 & -0.8 & -4.1 & -0.1 & -1.2 & 3.3 \\
XGM-SGG & -0.6 & -1.5 &-2.3 & 0.9 &-0.6 & 1.9 \\
\end{tabular}
\end{ruledtabular}
\end{table}

\subsection{Gravity field models for the Moon}
\begin{figure}
\captionsetup{font=small}
\includegraphics[width=\columnwidth]{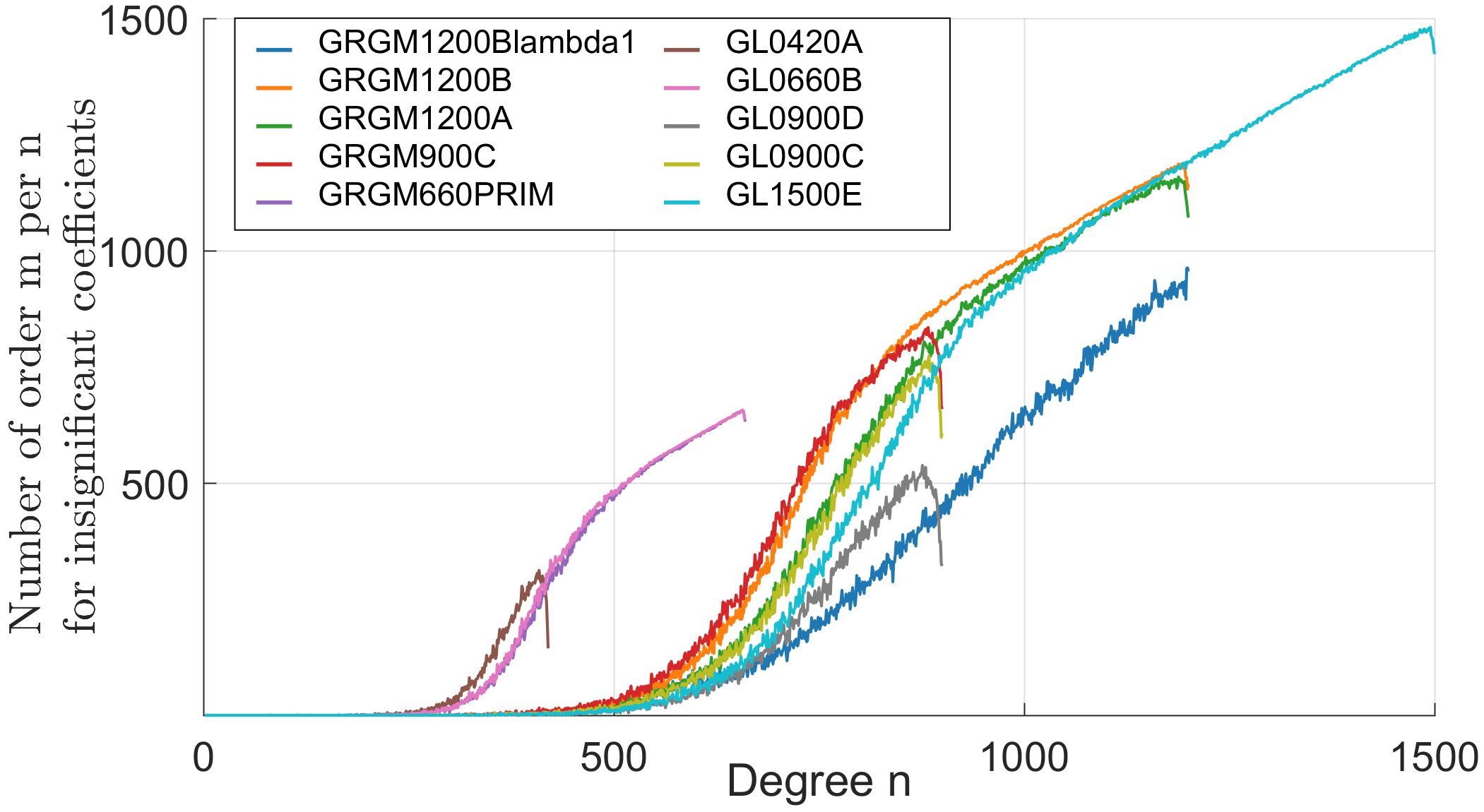}
\caption{\label{Fig5}Comparison of LGFMs.}
\end{figure}

Ten lunar gravity field models (LGFMs) \cite{Kahan2022} (names in Tab.~\ref{Tab6}, 1st column) were compared for $V_{L\text{st},\ell}/c^{2}$ at five L-clocks. Fig.~\ref{Fig5} highlights, for each degree $n$ in each model, the number of orders $m$ with insignificant normalized Stokes coefficients, defined by $\delta\bar{C}_{nm} > |\bar{C}_{nm}|$ or $\delta\bar{S}_{nm} > |\bar{S}_{nm}|$ or both, with uncertainties $\delta\bar{C}_{nm}$ and $\delta\bar{S}_{nm}$. For each model, beyond a threshold degree $n_{thr}$ (Tab.~\ref{Tab6}, 2nd column), the number of statistically insignificant coefficients rises sharply, indicating a noise-dominated tail. To quantify this behavior, let $N_{\rm insig}(n)$ denote the number of statistically insignificant orders at degree $n$. Since degree $n$ contains $n+1$ orders, for a degree interval $I$ we define the fraction of statistically insignificant orders as  %$P_I=\sum_{n\in I}N_{\rm insig}(n)/\sum_{n\in I}(n+1)$. 
\[P_I =\frac{\displaystyle\sum_{n\in I} N_{\mathrm{insig}}(n)}{\displaystyle\sum_{n\in I}(n+1)}.\]
For the retained degree ranges, $2\leq n\leq n_{\rm thr}$, this fraction is only $0.39\%$--$1.54\%$ across the ten models. In contrast, within only the first 50 degrees above $n_{\rm thr}$, the fraction already reaches $3.07\%$--$12.38\%$, and in the following 50-degree interval it increases further to $5.96\%$--$38.35\%$. These results quantitatively characterize the transition shown in Fig.~\ref{Fig5} and support the adopted $n_{\rm thr}$ as a conservative cutoff before the increasingly noise-dominated high-degree part of each model. Tab.~\ref{Tab6} reports $V_{Lst,\ell}/c^{2}$ uncertainties from the degrees $n\le n_{thr}$ of ten models. Tab.~\ref{Tab7} shows inter-model differences in $V_{Lst,\ell}/c^{2}$ among GL0900D, GL0900C and GRGM900C, with the same max model degree ($n_{mod}=900$) and threshold degree ($n_{thr}=500$) for a fairer comparison. Each model pair is compared using coefficients up to $n_{thr}$ and $n_{mod}$.   
\begin{table}
\captionsetup{font=small}
\caption{\label{Tab6}$V_{L\text{st},\ell}/c^{2}$ uncertainties from degrees $n \le n_{\mathrm{thr}}$ at L-clocks using ten LGFMs. Noise-dominated for $n>n_{\mathrm{thr}}$.}
\begin{ruledtabular}
\begin{tabular}{lc|ccccc}
LGFM & $n_{\mathrm{thr}}$ & A11 & L21 & A14 & A15 & L17 \\
\multicolumn{2}{c|}{} & \multicolumn{5}{c}{\scriptsize all entries $\times 10^{-19}$} \\
\midrule
GL0420A \cite{Konopliv2013}   & 250 & 2.7 & 3.1 & 2.4 & 2.8 & 3.1 \\
GL0660B \cite{Konopliv2013}  & 300 & 4.7 & 5.2 & 4.2 & 4.6 & 5.1 \\
GRGM660PRIM \cite{Lemoine2013} & 300 & 5.2 & 5.7 & 4.5 & 5.0 & 5.7 \\
GRGM900C \cite{Lemoine2014}& 500 & 6.0 & 6.6 & 4.8 & 5.3 & 5.4 \\
GL0900C \cite{Konopliv2014}               & 500 & 3.1 & 3.5 & 2.6 & 2.8 & 2.8 \\
\textbf{GL0900D} \cite{Konopliv2014}      & \textbf{500} & \textbf{1.9} & \textbf{2.1} & \textbf{1.5} & \textbf{1.7} & \textbf{1.7} \\
GRGM1200A \cite{Lemoine2014}& 500 & 3.0 & 3.4 & 2.4 & 2.9 & 2.8 \\
GRGM1200B \cite{Goossens2020}& 500 & 3.0 & 3.4 & 2.4 & 2.9 & 3.1 \\
GRGM1200Blambda1 \cite{Goossens2020}& 500 & 3.0 & 3.4 & 2.4 & 2.9 & 3.1 \\
GL1500E \cite{Konopliv2014} & 500 & 2.1 & 2.3 & 1.7 & 1.8 & 1.8 \\
\end{tabular}
\end{ruledtabular}
\end{table}

\begin{table}
\captionsetup{font=small}
\caption{\label{Tab7}Inter-model differences in $V_{L\text{st},\ell}/c^{2}$ at L-clocks (m1=GL0900D, m2=GL0900C, m3=GRGM900C).}
\begin{ruledtabular}
\begin{tabular}{l|cc|cc|cc}
\multicolumn{1}{c|}{L-clock} &
\multicolumn{2}{c|}{m1-m2 $(\times 10^{-18})$}&
\multicolumn{2}{c|}{m1-m3 $(\times 10^{-18})$}&
\multicolumn{2}{c}{m2-m3 $(\times 10^{-18})$} \\
& $n\le500$ & $n\le900$ & $n\le500$ & $n\le900$ & $n\le500$ & $n\le900$ \\
\midrule
A11 & 0.3 & 1.5   & 2.3  & 3.9   & 1.9  & 2.3   \\
L21 & 0.2 & 0.4   & -0.4 & -5.4 & -0.5 & -5.8 \\
A14 & 0.4 & -2.6 & 1.4  & -11.5 & 1.0  & -8.8 \\
A15 & 0.3 & 18.0  & 1.9  & 4.1   & 1.6  & -13.9 \\
L17 & 0.9 & 21.3  & 0.8  & 38.8  & -0.1 & 17.5 \\
\end{tabular}
\end{ruledtabular}
\end{table}

\begin{figure}
\captionsetup{font=small}
\includegraphics[width=\columnwidth]{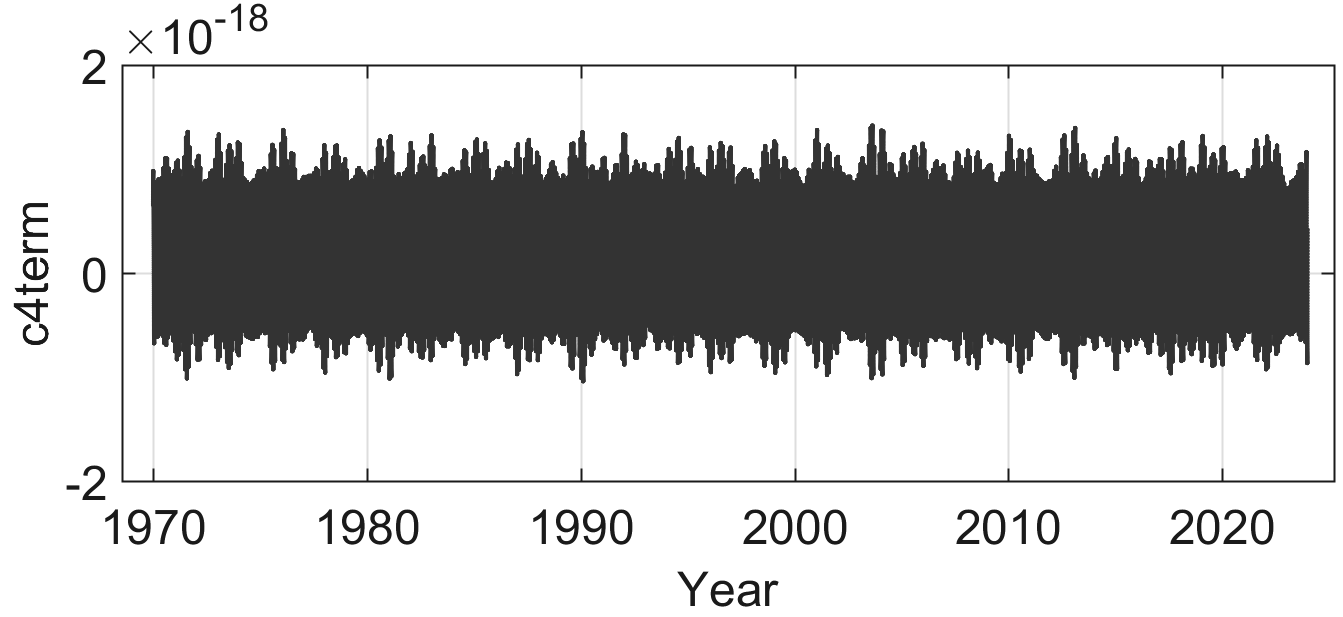}
\caption{Effect of $c^{-4}$ terms of $F/c^2$ in Eq.~\eqref{eq8}.}
\label{Fig6}
\end{figure}

\begin{figure}
\captionsetup{font=small}
\includegraphics[width=\columnwidth]{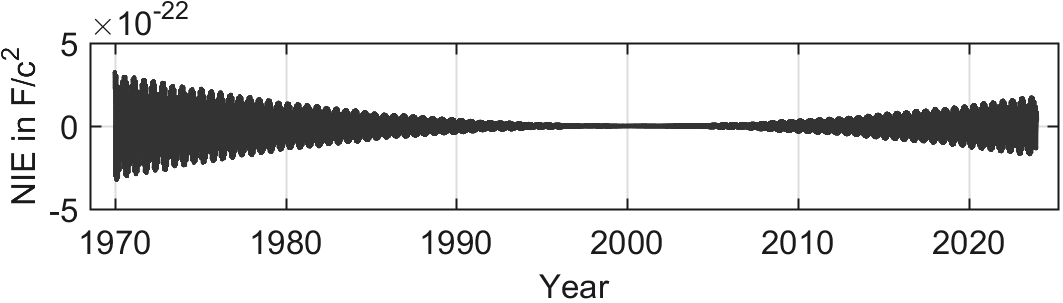}
\caption{\label{Fig7}Numerical integration error (NIE) of $F/c^2$.}
\end{figure} 
%%%%%%%%%%%%%Appendix C%%%%%%%%%%%%%%%%%%%
\section{Further remarks on $F/c^2$ in Eq.~\eqref{eq8}} 
\label{appC} 
\begin{table*}
\captionsetup{font=small}    
\caption{\label{Tab8}Uncertainty of $F/c^2$ in Eq.~\eqref{eq8} caused by the adopted body position accuracies. Inner planets denote Mercury, Venus, Earth and Mars. The range $[0.6,\,5.5]\times10^{-22}$ is the combined propagated uncertainty from all non-Sun bodies listed in the table, whereas $[1.1,\,2.6]\times10^{-19}$ is the propagated uncertainty from the Sun.}
\begin{ruledtabular}
\begin{tabular}{cccc}
\shortstack[c]{Body\\ {}} &
\shortstack[c]{Given position accuracy\\ in \cite{Folkner2014}} &
\shortstack[c]{Adopted position accuracy\\ in calculation [km]} &
\shortstack[c]{$F/c^{2}$ uncertainty\\from adopted accuracy} \\
    \midrule
   \text{Moon (w.r.t. Earth)}
 & sub-meter & $10^{-3}$ & \multirow[c]{4}{*}{\shortstack[c]{All non-Sun bodies combined\\$[0.6,\,5.5]\times10^{-22}$}} \\
    Inner planets & sub-km &1.0 &\\
    Jupiter, Saturn & tens of km & $10^{2}$ &\\    
    Uranus, Neptune &several thousand km& $10^{4}$ &\\
    \midrule
    Sun & n/a & 0.7 & $[1.1,2.6]\times10^{-19}$ \\
\end{tabular}
\end{ruledtabular}
\end{table*}
\subsection{$c^{-4}$ terms in $F/c^2$}
Based on \cite[Eq.~(58)]{Soffel2003}, we derive $c^{-4}$ terms of $F/c^2$ in Eq.~\eqref{eq8}:
\begin{equation}
\begin{aligned}
\delta_F^{(4)}=\frac{1}{c^4}\Bigg[
&\frac{v_E^4-v_L^4}{8}
+\frac{3}{2}(v_E^2U_E^{\mathrm{ext}}-v_L^2U_L^{\mathrm{ext}})\\
&-\frac{(U_E^{\mathrm{ext}})^2-(U_L^{\mathrm{ext}})^2}{2}\\
&-4(U_E^{\mathrm{ext}}\bm{v}_E\cdot\bm{v}_A
-U_L^{\mathrm{ext}}\bm{v}_L\cdot\bm{v}_A)\\
&+(\bm{v}_L\cdot\bm{a}_L-3\dot{U}_L^{\mathrm{ext}})
\bm{v}_L\cdot\bm{r}_{\ell L}\\
&-(\bm{v}_E\cdot\bm{a}_E-3\dot{U}_E^{\mathrm{ext}})
\bm{v}_E\cdot\bm{r}_{\ell E}\\
&+\left(3U_E^{\mathrm{ext}}+\frac{v_E^2}{2}\right)
(\bm{a}_E\cdot\bm{r}_{\ell E}
+\bm{v}_E\cdot\bm{v}_{\ell E})\\
&-\left(3U_L^{\mathrm{ext}}+\frac{v_L^2}{2}\right)
(\bm{a}_L\cdot\bm{r}_{\ell L}
+\bm{v}_L\cdot\bm{v}_{\ell L})
\Bigg].
\end{aligned}
\label{eq:delta_F4}
\end{equation}
In the BCRS, $\bm{v}_E$ and $\bm{v}_L$ are the barycentric velocities of the Earth ($E$) and Moon ($L$), and $\bm{a}_E$, $\bm{a}_L$ their accelerations. The relative positions $\bm{r}_{\ell L}$ and $\bm{r}_{\ell E}$ denote the L-clock $\ell$ with respect to the Moon and Earth, with corresponding relative velocities $\bm{v}_{\ell L}$ and $\bm{v}_{\ell E}$. The external Newtonian potential and its time derivative at body $B\in\{E,L\}$ are 
\[
\begin{aligned}
U_B^{\mathrm{ext}}
&=\sum_{A\ne B}\frac{GM_A}{r_{BA}},\\
\dot{U}_B^{\mathrm{ext}}
&=-\sum_{A\ne B}
\frac{GM_A\,\bm{r}_{BA}\cdot\bm{v}_{BA}}{r_{BA}^{3}}.
\end{aligned}
\]
$\bm{v}_A$ is the velocity of body $A$. As shown in Fig.~\ref{Fig6}, the value range of $\delta_F^{(4)}$ is $[-1.0,\,1.4]\times 10^{-18}$ for all the assumed five L-clocks.

\subsection{Uncertainty in $F/c^2$ from celestial body position accuracy and numerical integration}
Tab.~\ref{Tab8} (second column) lists the Moon and planet position accuracies from \cite{Folkner2014}. The more conservative accuracies (i.e., worse accuracies) adopted in this work are given in the third column and are used to estimate their propagated contributions to the uncertainty of $F/c^2$ in Eq.~\eqref{eq8}, as shown in the fourth column. The Sun's position accuracy is not given directly in \cite{Folkner2014}; the adopted value in the third column is derived from the accuracies of the other bodies via \cite[Eq.~(4)]{Folkner2014}. Even with these conservative accuracies, the uncertainty in $F/c^2$ remains below $10^{-18}$: it is dominated by the Sun with $[1.1,2.6]\times10^{-19}$, while the contributions from the other bodies are only $[0.6,5.5]\times10^{-22}$. Fig.~\ref{Fig7} shows that the numerical integration errors in $F/c^2$ of Eq.~\eqref{eq8} over 50 years vary within $[-3.2, 3.2]\times10^{-22}$, far below $10^{-18}$.
%%%%%%%%%%%%%%%%References%%%%%%%%%%%%%%%%%%%%%%%
\clearpage
\bibliography{References}
\end{document}